%% file: 0.main.tex
\documentclass[english,11pt]{article}

\usepackage{fullpage} % Sets small margins
\usepackage[section]{placeins} % put a barrier so floats (figs) stay within a section
\usepackage[doublespacing]{setspace}

\usepackage{amsmath} % For equations and other math symbols
\usepackage{lmodern}
\usepackage[T1]{fontenc}
\usepackage[latin9]{inputenc}
\usepackage{amssymb} %For checkmark symbols

\usepackage{booktabs} %For nice tables
\usepackage{caption} %Captions and sub captions in tables and figures
\usepackage{subcaption} %Subfigures and sub-captions
\usepackage{pdflscape} % For rotating tables
\usepackage[flushleft]{threeparttable} %For neat notes below tables
\usepackage{array}

\usepackage{graphicx}
\usepackage[dvipsnames]{xcolor} %Use xcolor palette

\usepackage[colorlinks]{hyperref} %Citations and references
    \hypersetup{colorlinks,citecolor=MidnightBlue, linkcolor=black, urlcolor=blue}
\usepackage[sort&compress,comma,authoryear]{natbib}

\usepackage{soul} %For highlighting text with the \hl command
\usepackage{todonotes} %Add "to do" notes on the side of the document

\usepackage{tikz}
    \usetikzlibrary{arrows,positioning,decorations.pathreplacing}

\usepackage{appendix}
\newcommand{\ignore}[1]{}

\usepackage{caption}
\begin{document}
\title{Community Matters: \\ Heterogeneous Impacts of a Sanitation Intervention\thanks{%
This research was carried out as part of the Formal Research Component of WaterAid UK's Project "Sustainable Total Sanitation Nigeria -implementation, learning, research, and influence on practice and policy'' (STS Nigeria), funded by the Bill and Melinda Gates Foundation. Abramovsky, Augsburg and Oteiza are also grateful for the funding from the ESRC-funded Centre for the Microeconomic Analysis of Public Policy (CPP) at IFS, grant reference ES/M010147/1. The evaluation is registered at the ISRCTN registry (ISRCTN74165567). The authors would like to thank the participants of the EDePo Development Workshop, 3ie seminar, ESPE, Royal Economic Society conference and the London School of Hygiene and Tropical Medicine for helpful comments and suggestions. Corresponding author Britta Augsburg, The Institute for Fiscal Studies, 7 Ridgmount Street, London WC1E 7AE; Phone: 0044(0)20 7291 4800, Fax: 0044(0)20 7323 4780, Email: britta\_a@ifs.org.uk.}}
\begin{singlespace}
\author{Laura Abramovsky \and Britta Augsburg \and Melanie L\"uhrmann  \and Francisco Oteiza \and Juan Pablo Rud}
\end{singlespace}
\date{}
%\author{Laura Abramovsky\thanks{Centre for the Evaluation of Social Policies (EDePo), Institute for Fiscal Studies} \and Britta Augsburg\footnotemark[2] \and Melanie L\"uhrmann\thanks{Department of Economics, Royal Holloway, and IFS} \and Francisco Oteiza\thanks{Institute of Education, University College London} \and Juan Pablo Rud\footnotemark[3]}

%\date{This draft: May 2019}

\maketitle

%%%%%%%%%%%%%%%%%
% ABSTRACT
%%%%%%%%%%%%%%%%%

\begin{singlespace}

%\begin{} 

%We study the effectiveness of a popular community-level information intervention aimed at improving sanitation in low income countries in a cluster-randomized controlled trial (RCT) in 247 Nigerian communities. We evaluate the intervention, Community-Led Total Sanitation (CLTS),which is part of national sanitation policy in more than 25 countries, at scale. While average impacts almost three years after implementation are exiguous, the results hide important heterogeneity. The intervention has strong and lasting effects on sanitation practices in poorer communities. These are realized through sanitation investments following the intervention. Our results are robust across alternative measures of communities’ socio-economic characteristics. We identify community wealth as key statistic for effective intervention targeting that is widely available in secondary data. Finally,we show that community-level wealth heterogeneity across these trials can rationalize the wide rangeof impact estimates found in five other CLTS RCTs across various contexts, providing plausibleexternal validity to our findings with implications for intervention scale-up

We study the effectiveness of a participatory community-level information intervention aimed at improving sanitation using a cluster-randomized controlled trial (RCT) in Nigerian communities. The intervention, Community-Led Total Sanitation (CLTS), is currently part of national sanitation policy in more than 25 countries. While average impacts are exiguous almost three years after implementation at scale, the results hide important heterogeneity: the intervention has strong and lasting effects on sanitation practices in poorer communities. These are realized through increased sanitation investments. We show that community wealth, widely available in secondary data, is a key statistic for effective intervention targeting. Using data from five other similar randomized interventions in various contexts, we find that community-level wealth heterogeneity can rationalize the wide range of impact estimates in the literature. This exercise provides plausible external validity to our findings, with implications for intervention scale-up.\\ 
\emph{JEL Codes:} O12, I12, I15, I18.

%\end{}

\noindent 

%\noindent \emph{Keywords:} External validity, Heterogeneous Treatment Effects, Sanitation, Information, Cluster-Randomized Control Trial.

\end{singlespace}

\newpage

%%%%%%%%%%%%%%%%%
% BODY SECTIONS
%%%%%%%%%%%%%%%%%
%\input{1.introduction.tex} 
\input{1.introductionv2.tex}

\input{2.clts.tex} 
\input{3.rdesign.tex}

\input{5.IEresults_new.tex}

\input{6.comparisons.tex} 
\input{8.conclusion.tex}

%%%%%%%%%%%%%%%%%
% BIBLIOGRAPHY
%%%%%%%%%%%%%%%%%
\newpage
\bibliography{refs}
	\bibliographystyle{apa}

%%%%%%%%%%%%%%%%%
% APPENDICES
%%%%%%%%%%%%%%%%%
\appendix
\input{9a.Appendix.tex}

\input{9b.Appendix.tex}

\end{document}

%% file: 1.introductionv2.tex
\section{Introduction}\label{sec:intro}
Public health interventions %information campaigns 
are often promoted as an instrumental driver of behavioral change and  adoption of health products. But how effective are these programs in developing countries, especially when implemented at scale?  We investigate this question in the context of a participatory community-level intervention that has been widely introduced by governments around the world to improve access to safe sanitation. 

Comprehensive water and sanitation programmes in developed nations in the early 1900s have been dubbed the most effective public health intervention of the last century \citep{AlsanGoldin2019} due to their significant impact on infant survival and reduction of communicable diseases. The costs of poor sanitation and the disease environments they create - in terms of child health, mortality, human capital accumulation, and economic growth - are, by now, well understood \citep{Adukia2017,Alzua2018ASanitation,SpearsLamba2016,AugLes2018,Bairdetal2016}.\footnote{ Low and middle income countries carry a disproportionately heavy burden for communicable diseases, such as diarrhea \citep{PrussUstunCorvalan2006}. Better sanitation could, for example, prevent the majority of diarrhea-related deaths of 361,000 children aged less than 5 years each year \citep{pruss2014burden}, and would reduce the prevalence of worm infections in school children, which affect educational attainment and labor market outcomes \citep{Bairdetal2016}. For an overview, see also \citet{UnitedNations2016CleanMatters,WSP2012}} Yet, 4.5 billion people still lack access to safely managed sanitation worldwide \citep{WHOUNICEFJMP2017}. Improving access to sanitation has thus been recognized as a goal towards sustainable development by the UN. To achieve it, identifying effective policies that work in low income countries is key.

In this paper, we show the results of a cluster randomized controlled trial (RCT) we implemented to assess the Government of Nigeria's `National Strategy for Scaling up Sanitation and Hygiene'.\footnote{ Nigeria provides a suitable context to study sanitation as 34\% of its population practices open defecation, toilet ownership rates have stagnated \citep{unicef2015} and the country contributes to a significant share of the global population without access to adequate sanitation \citep{WHOUNICEFJMP2017}.} We evaluate the main pillar of the governmental strategy, an intervention known as Community-Led Total Sanitation (CLTS), at scale. CLTS entails community meetings and the provision of information with the aim of eradicating open defecation (OD), by triggering a desire for collective behavioral change and encouraging communities to construct and use toilets. As part of the experiment, CLTS was implemented in a random sample of 125 of 247 study clusters of rural communities, located in the Nigerian states Ekiti and Enugu.
In each cluster, we randomly selected 20 households for interview at baseline in 2014 and conducted three follow-up surveys with this sample 8, 24 and 32 months after implementation. The resulting balanced panel contains more than 4,500 households.

% new: exchanged the first sentence below with this one: (BRITTA_26June: Slight adjustment to sentence made)
Our results show that, on average, CLTS led to small and temporary reductions in OD among households in treated communities, i.e. a reduction by 3 percentage points (pp) over a period of 24 months. However, these estimates hide meaningful impact heterogeneity across population subgroups: we find that intervention impacts are considerably stronger among, and restricted to, the asset poorest half of the studied communities. In these communities, OD fell by three times as much as on average, i.e. 9 percentage points in the short-run (8 months post implementation), and this sanitation improvement \emph{is} sustained over time. The reduction in OD is achieved mainly through household sanitation investments, i.e. the construction of new toilets. The effect cannot be ascribed to pre-treatment differences between communities in toilet ownership. Neither do we find evidence that community differences in social capital or social interactions, public goods infrastructure or leader characteristics explain our results. Yet, communities' socio-economic status clearly matters for the intervention effectiveness. Community-specific impact estimates along  three alternative socio-economic dimensions of communities - geographic isolation, low population density and low average night light intensity (obtained from satellite measurements) - yield estimates of similar magnitude to those for community wealth.

Additionally we use results from the Nigerian RCT and data from five recently published CLTS trials conducted in other countries \citep{bricenoEtAl2017,Patil2014TheTrial,Cameron2019,pickering2015effect,guiteras2015encouraging} to i) demonstrate that such heterogeneous impacts can also be found in other contexts, and to ii) show an inverse relationship between area-level wealth and program effectiveness \emph{beyond the Nigerian context}. This provides external validity to our findings and rationalizes CLTS impact estimates across studies.

Our results have four important implications.
First, we show that public health information interventions such as CLTS can indeed trigger behavioral change in (some) population groups, in line with results surveyed in \citet{dupas2011health}. Second, and consistent with CLTS practitioners' experiences and priors as to where the intervention should work best \citep{kar2008handbook}, we pinpoint community wealth as a key factor for the effectiveness of this participatory community-level information intervention.\footnote{Recent studies have tried to identify the underlying factors that drive heterogeneous impacts of interventions across and within studies, such as \citet{Meager2018UnderstandingExperiments,Bandiera2018SocialServices, cunha2018price}.} 
%\citet{cunha2018price}, for example, investigate the role of community characteristics in shaping the effects of cash and in-kind transfers. They find that the degree of integration and the size of the markets explained differential effects in local prices.}  
We show that heterogeneous impacts by communities' socio-economic characteristics dominate impact heterogeneity at the household level.

Third, our results have important implications for the scale-up of CLTS. To be able to draw policy relevant conclusions regarding a wider rollout \citep{Muralidharan2017}, we implemented CLTS at scale. Government officials working in WASH units (Water, Sanitation and Hygiene) were trained to deliver the intervention. We argue, based on our findings, that a targeted implementation of CLTS may yield larger policy returns than a blanket approach. Measures of wealth are readily available in standard household surveys, or alternatively nightlight intensity indices from open access satellite data. Governments can easily combine these data with findings from RCTs to develop a more effective targeting strategy for this popular sanitation program, which is currently implemented in more than 25 Latin American, Asian and African countries. As an example, we demonstrate this using the 2013 Nigerian Demographic and Health Survey (DHS), and show that the data can replicate the classification of geographical areas into poorer and (modestly) richer ones with a precision that is similar to the measures obtained from our primary study data.

Finally, we demonstrate how heterogeneous intervention impacts can help address the widely acknowledged concern that the high internal validity that characterises RCTs comes with important shortcomings in external validity \citep{Basu2014,DeatonCartwright2018,PetersEtAl2018}. \citet{WangEtAl2006} and \citet{Meager2018UnderstandingExperiments} emphasize context or location-specific factors as obstacles to external validity. We present local socio-economic status, particularly community wealth, as common underlying factor for CLTS effectiveness \emph{beyond the Nigerian setting of our RCT} and, in fact, across various diverse contexts. The results from five recently published CLTS trials \citep{bricenoEtAl2017,Patil2014TheTrial,Cameron2019,pickering2015effect,guiteras2015encouraging} range from very large impacts - an increase in toilet ownership of 30 percentage points, and an open defecation reduction of 23 percentage points - in a trial in Mali \citep{pickering2015effect}, to no detectable impacts on toilet ownership in Bangladesh \citep{guiteras2015encouraging}. We estimate heterogeneous impacts for each of these evaluations using a harmonised method and pair impact estimates with a proxy of local socio-economic characteristics available for all studies, namely average night light intensity at baseline. The inverse relationship between area-level wealth and program effectiveness that we find in this exercise rationalizes the existing range of CLTS impact estimates across studies. We conclude, in the spirit of \citet{BanerjeeEtAl2017Chapter4}, that our result from Nigeria carries plausible external validity across various contexts. 

%We re-estimate impacts for each of these evaluations using a harmonised method and pair impact estimates with a proxy of local socio-economic characteristics available for all studies, namely the average night light intensity, observed at baseline. The inverse relationship between area-level wealth and program effectiveness that we find in this exercise rationalizes the existing range of CLTS impact estimates across studies. We conclude, in the spirit of \citet{BanerjeeEtAl2017Chapter4}, that our result from Nigeria carries plausible external validity across various contexts. 

The remainder of the paper is structured as follows. In the next section we describe the intervention and the experimental design. In Section 3, we present the empirical method and Section 4 presents our impact estimates. Section 5 compares the results of our study to those of other CLTS interventions. Section 6 concludes.

%We find heterogeneous impact estimates of similar magnitude for three alternative measures of low communities' socio-economic status, i.e. geographic isolation, low population density and low average night light intensity (obtained from satellite measurements). They cannot be ascribed to differences between communities in toilet ownership at baseline. We also do not find evidence that community differences in social capital or social interactions, public goods infrastructure or leader characteristics explain our results. 

% In a review study, \citet{CambonEtAl2012} identify ‘transferability’ as "a major limitation in the use of research results by health stakeholders and decision-makers, and thus in the process of evidence-based health education and promotion \citep{JuneauEtAl2011}". CLTS constitutes an apt example. First developed in Bangladesh in 1999, it is today a widely adopted intervention, implemented in more than 25 Latin American, Asian and African countries – despite available evidence on its effectiveness providing mixed results \citep{Garn2017TheMeta-analysis}, and judged to be weak and of limited credibility \citep{Venkataramanan2018}. 

%% file: 2.clts.tex
\section{Intervention and study design}\label{sec:clts}

\subsection{Community-Led Total Sanitation}\label{subsec:clts}

The Government of Nigeria adopted Community-Led Total Sanitation as its major approach for the development of rural sanitation within its Strategy for `Scaling up Sanitation and Hygiene', launched in 2007. This decision followed three years of piloting, conducted by the National Task Group for Sanitation in collaboration with state and local governments as well as local and international NGOs such as WaterAid and UNICEF. The effort to scale-up CLTS to the whole country began in 2008. We study the effectiveness of CLTS in the context of Nigeria's national strategy through a cluster-randomized controlled trial conducted in nine local government areas (LGAs) in the states of Enugu and Ekiti that did not have any recent experience of CLTS, or CLTS-like interventions.

CLTS is a community-level information and mobilization intervention aimed at reducing open defecation and improving toilet coverage. It is typically implemented in three steps. The firsts step focuses on mobilization: Community leaders are approached and engaged in a discussion about the negative health implications of OD\footnote{ A further key message is the importance of sanitation externalities, i.e. that all community members (particularly children) are at risk of contracting sanitation-related diseases if some residents practise open defecation.}, as well as the potential benefits of CLTS in achieving behavioral change within their communities. The aim of the meeting is to convince community leaders to arrange a community meeting. This meeting, the so-called `triggering meeting', marks the second step, and the main component of CLTS. The meeting starts once and only if a significant number of community members gathered in a predefined public space on the identified day. The first activity is typically a community mapping exercise, in which each attending community member marks their household's location and toilet ownership status on a stylized map on the ground. Community members next identify regular OD sites and mark these as well. In many cases, this exercise is used by facilitators to follow up with graphic images showing that the community lives in an environment contaminated by feces. Facilitators of the meetings further use the map to trace the community's contamination paths of human feces into water supplies and food.\footnote{ A number of other activities may follow, at the discretion of the facilitator. Examples include transect walks through the community (often referred to as `walks of shame'), pointing out visible feces in the environment to evoke further disgust and shame; medical expense calculations related to illnesses likely induced by OD practices; or graphic exercises, where facilitators might add feces to drinking water, illustrating that these are not necessarily visible to the naked eye. In the context of our study, only about 20\% of triggering meetings included at least one such additional exercise, graphic illustration being the most popular one (implemented in 14\% of triggering meetings) followed by expense calculations (7\%).}

As a closing task, attendees are asked to draw up a community action plan to achieve community-level open defecation free (ODF) status. This aspect of CLTS seeks to foster collective action and collaboration. It includes discussions of how poor or vulnerable households can be supported to achieve ODF status. The action plan is posted in a public spot. Volunteers (so-called `natural leaders') are chosen to follow up regularly on each attendee's commitment towards implementing the plan, the thirs step of CLTS. Followup visits by the facilitators are organized. Eventually, the community might be certified for its achievements by the national Rural Water Supply and Sanitation Agency (RWASSA) and the National Task Group on Sanitation (NTGS).

CLTS does not offer any monetary incentives, subsidies or credit to finance toilet construction or reward OD reductions or ODF achievement, nor is technical assistance or hardware provided. Neither does it promote a particular toilet technology. The aim is to drive a change in sanitation practices purely by altering the perceived costs of unsafe sanitation and the perceived benefits of toilet use.

%% file: 3.rdesign.tex
%\section{The Experiment}\label{sec:experiment}

\subsection{Experimental design and intervention implementation}\label{sec:design}

\begin{figure}[htbp]
	\centering
	\caption{Map of the study area in the Nigerian states of Ekiti and Enugu}\label{f:map}
		\includegraphics[width=\textwidth]{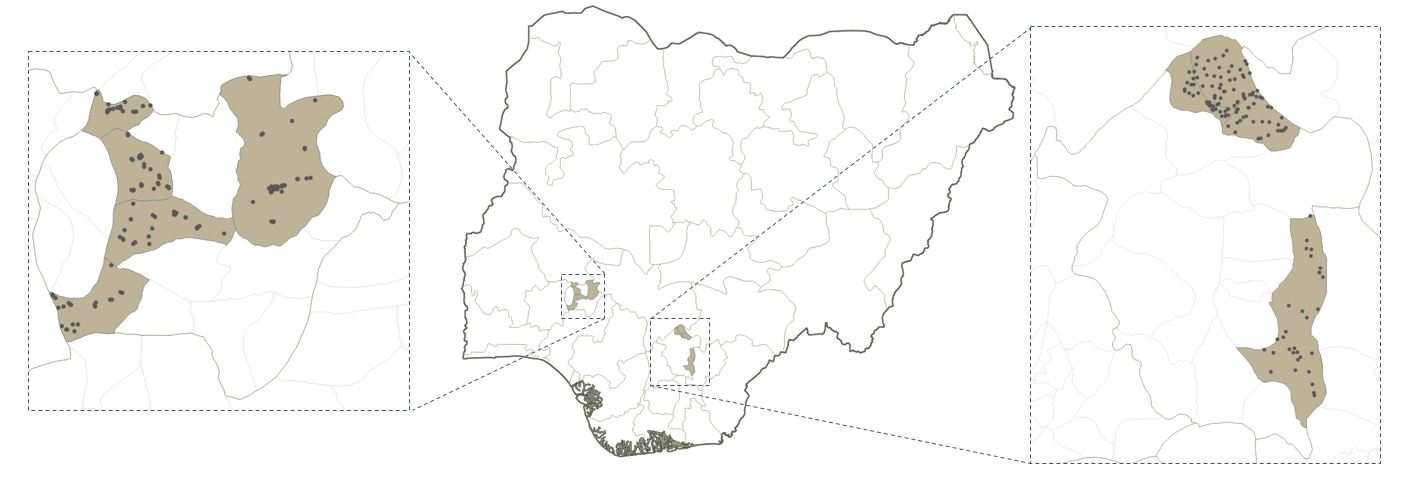}
		\caption*{\emph{Note:} Study clusters in the nine selected LGAs (shaded) from Ekiti (left) and Enugu (right).}
		\label{Locations}
\end{figure}

We study the effectiveness of CLTS in the context of Nigeria's national strategy through an RCT conducted in collaboration with the international non-governmental organisation (NGO) WaterAid.\footnote{ The study protocol was approved by the following IRBs: National Health Research Ethics Committee, Federal Ministry of Health, Nigeria (NHREC/01/01/2007-20-20/11/2014), University College London Ethics Committee (2168/009). The trial was registered at the ISRCTN registry (ISRCTN74165567). We note that the research project intended to also evaluate a supply intervention, Sanitation Marketing (SanMark). However, SanMark development and piloting took longer than planned and implementation had been in place not long enough at the time of the endline survey to conduct a full impact analysis. Details are outlined in the project's final report \citep{AbramovskyEtAl2018}.} WaterAid worked closely with local government areas (LGAs)\footnote{ LGAs are Nigeria's second sub-division after states. LGAs are administrative divisions led by a Local Government Council.} and two local resource agencies (NGOs from Ekiti and Enugu) to implement CLTS in two of Nigeria's 36 states, Ekiti, Enugu.\footnote{ A third state, Jigawa, was dropped due to budget limitations and security concerns.} Specifically, WaterAid Nigeria and the two local NGOs conducted CLTS training sessions, one in each state. These sessions trained the LGA water, sanitation and hygiene (WASH) units, which are part of Nigeria's public service, who then conducted the CLTS mobilization, triggering meetings and followup activities. Four LGAs without recent experience of CLTS, or CLTS-like interventions, were selected in Enugu and 5 in Ekiti. Their locations are indicated in Figure~\ref{Locations}.\footnote{ Study LGAs in Enugu are Igbo Eze North, Igbo Eze South, Nkanu East and Udenu. In Ekiti, Ido Osi, Ikole, Moba, Irepodun Ifelodun and Ekiti South West are part of the study.}

% Enugu and Ekiti Figure

Since CLTS is a community-level intervention, we used what we refer to as `triggerable clusters' as the unit of randomization. Triggerable clusters are groups of geographically close villages, neighbourhoods or quarters. These clusters were defined jointly by the researchers and the implementers. Hence, they do not match any of Nigeria's administrative units. Clusters were chosen with the view of reducing information spillover, e.g. to be self-contained units so that information about triggering activities would not spread to the next cluster. This implies, for example, that triggerable clusters do not share markets or large public areas. %should we explain that villages are in Enugu and neighbourhoods/quarters in Ekiti? Maybe in footnote?
To further reduce potential contamination between experimental treatment and control clusters, `buffer' areas were introduced around triggerable clusters to ensure that no two clusters were located in close geographic proximity. No specific distance was imposed. The definition of the `triggerable clusters' and their buffer zones was driven by implementers' previous experiences from working in these areas.

In both study states, a cluster comprises of on average 1.7 villages or quarters\footnote{ The median and modal number of villages or quarters within a cluster is 1. The maximum number of villages in a cluster is 7, occurring only once.}, all of which CLTS was implemented at the same time. The treatment period during which mobilization and triggering activities took place lasted about six months -- between January and June 2015.

In total, we identified 247 `triggerable clusters'. 246 of these were randomized with equal probability into either receiving CLTS (treatment) or not receiving it during the course of the study (control).\footnote{ One cluster from the original sampling was subsequently dropped since data collection in any post-treatment survey wave was not possible due to civil unrest in the community.} Randomization was stratified by LGA. The distribution of treatment and control clusters is presented in Table~\ref{RandomizationLGAs}. 

\begin{table}[ht]
    \centering 
    \caption{Number of triggerable clusters per study arm and state}   
    \footnotesize 
    \begin{tabular}{l c c c c c c }  
    \toprule
    & \multicolumn{2}{c}{Control} & \multicolumn{2}{c}{CLTS} & \multicolumn{2}{c}{Total} \\
    & Freq. & \%  & Freq. & \% & Freq. & \% \\
    \midrule
    Ekiti & 63 & 51.6 & 66 & 52.8 & 129 & 52.2 \\
    Enugu & 59 & 48.4 & 59 & 47.2 & 118 & 47.8  \\
    Total &  122 & 100.0 & 125 & 100.0 & 247 & 100.0 \\
    \bottomrule
    \end{tabular}%   
    \label{RandomizationLGAs}
\end{table}

\subsection{Respondent sampling, data collection and attrition}\label{sec:sample}

We collected five rounds of data over a time span of almost three and a half years.\footnote{Data collection was carried out by an independent data collection company, blinded to treatment status.} The sampling frame was established in October 2014 through the first round of data collection, a household census in the nine participating LGAs from Ekiti and Enugu. The census collected basic household information from 50,333 households (27,888 from Enugu and 22,445 from Ekiti).\footnote{ Initial estimates of the size of triggering clusters in the implementation plan were around 150 households. In the field, cluster sizes turned out to be larger. Budget limitations constrained the listing to a maximum of 180 households per cluster. To achieve a representative sample of households, we adopted the following approach: For each triggerable cluster, we randomly ordered the villages/quarters and started the listing exercise at the top of the list. The first village/quarter was listed completely, independent of its size (i.e. going above the 180 households threshold if needed). If less than 150 households were listed, data collection would continue in the second village/quarter on the list, again listing every household in this village/quarter. This process continued until either all villages/quarters were listed in each cluster, or until around 180 households were reached. This approach ensured that we have listing data from each of the study clusters, that our overall sample remains representative for the study area (since the ordering of village/quarter listing was randomly determined), and furthermore that whole villages and quarters were listed while keeping within budget.}
Based on this census, we randomly selected 20 households from each cluster for interview. Our sample is thus a representative panel of households in the nine LGAs. This is in contrast to other studies which restricted their samples to households with children (\cite{Cameron2019,briceno2015promoting,pickering2015effect}). Our final sample consists of 4,671 households in the 246 clusters, distributed evenly across Ekiti and Enugu, and covering around 9\% of the population in the study area.
A baseline survey was conducted during December 2014 and January 2015. The first follow-up survey (FU1) took place between December 2015 and February 2016 eight months after implementation, on average. FU2 took place 24 months after implementation (March-April 2017) and FU3 (which we also refer to as the `endline survey') after 32 months (between November 2017 and January 2018).
The three followup surveys allow us to study the dynamics of CLTS impacts over time, providing insight into the sustainability of program impacts. Figure \ref{f:timeline} summarizes intervention and data collection timings.

%%%%%%%%%%%%%%%%%%%%%%%%%%%%%%%%%%%%%%%%%%%%%%%%
% TIMELINE (TIKZ VERSION, MORE PARSIMONIOUS)
\begin{figure}[!ht]
    \begin{center}
    \caption{Project timeline: implementation and data collection waves} \label{f:timeline}
        \begin{tikzpicture}[
            every node/.style = {align=center},
            Line/.style = {-angle 90, shorten >=2pt},
            Brace/.style args = {#1}{semithick, decorate, decoration={brace,#1,raise=2pt,
                             pre=moveto,pre length=2pt,post=moveto,post length=2pt,}},
            ys/.style = {yshift=#1}
            ]
            
\linespread{0.8}                    

% Set coordinate points of reference
\coordinate (a) at (0,0);
\coordinate[right=15mm of a]    (b);
\coordinate[right=15mm of b]    (c);
\coordinate[right=15mm of c]    (d);
\coordinate[right=15mm of d]    (e);
\coordinate[right=15mm of e]    (f);
\coordinate[right=15mm of f]    (g);
\coordinate[right=15mm of g]    (h);
\coordinate[right=15mm of h]    (i);
\coordinate[right=15mm of i]    (j);

% Draw the line
\draw[Line] (a) -- (j) node[below left] {time};
\draw[](15mm,5mm)node[font=\large]{2014};
\draw[](45mm,5mm)node[font=\large]{2015};
\draw[](75mm,5mm)node[font=\large]{2016};
\draw[](105mm,5mm)node[font=\large]{2017};

% Draw the vertical tick marks and nodes for each period of interest
\draw (0 mm,5pt) -- (0 mm,-5pt);
\draw (30 mm,5pt) -- (30 mm,-5pt);
\draw (60 mm,5pt) -- (60 mm,-5pt);
\draw (90 mm,5pt) -- (90 mm,-5pt);
\draw (120 mm,5pt) -- (120 mm,-5pt);

% colored bar down
\draw[lightgray, line width=8pt] 
(31mm,0pt) -- (44mm,0pt);

% Draw the rest of the features
\draw[Brace=mirror] (b) -- node[below=5pt] {Baseline} (c);
\draw[] (c) -- node[below=5pt] {CLTS} (d);
\draw[Brace=mirror] (e) -- node[below=5pt] {FU 1} (f);
\draw[Brace=mirror] (g) -- node[below=5pt] {FU 2} (h);
\draw[Brace=mirror] (h) -- node[below=5pt] {FU3} (i);
\end{tikzpicture}
\end{center}
\end{figure}

Our study had low attrition rates across the three followup survey rounds: 2.53\% at FU1, 8.81\% at FU2 and 11.58\% at FU3 as shown in Appendix Table~\ref{t:att1}. %The fact that 3 years after the baseline survey, we successfully interviewed almost 90\% of the households in our initial sample is encouraging. 
A possible concern is that non-response was not random across treatment and control villages, which could bias estimated treatment effects. The lower panel of Table \ref{t:balanceall} shows that there is no significant difference between attrition levels in the control and intervention clusters at the time of the endline survey, FU3, with attrition rates of 11.14 for households in control and 12.05 in treatment areas. The same conclusion holds for differential attrition by treatment status during FU1 and FU2 (Appendix Table \ref{t:att1}).
To investigate this more formally, we test whether treatment status can predict attrition conditional on district fixed effects and household-level baseline characteristics (Appendix Table \ref{t:att2}). Overall, the coefficient of the treatment dummy is close to zero and never statistically significant in any of the three followup surveys or estimated specifications. This suggests that selective attrition is not a concern. We therefore focus our analysis on the balanced panel of 4,540 households which were successfully interviewed in each round of data collection.

\subsection{The study population and treatment-control balance}\label{sec:outcomes}

As the allocation of clusters to treatment and control was random, we expect no systematic differences between both groups at the time of the baseline survey. We check balancedness in Table \ref{t:balanceall}, presenting summary statistics measured at baseline for the main characteristics of households, the outcomes we consider and characteristics of the study communities. We present the baseline mean for the control group (in the post attrition sample), the difference in means between treatment and control group and report the p-value for a t-test of equality of these means in the last column.

As expected, there are no statistically significant differences between the two groups along the presented dimensions,\footnote{ The project's baseline report supports balancedness on a wide range of additional variables \citep{Abramovsky2015BaselineNigeria}.} except a small (0.27) difference in the number of household members. In an F-test of joint significance of all characteristics, we reject the null hypothesis at the 5\% level (p-value=0.038), and hence include this variable as a covariate throughout our analysis. Yet, once we remove household size, the explanatory power of the remaining variables falls markedly (p-value=0.27), supporting the validity of our randomization strategy (except regarding household size).

\begin{table}
	\centering
	\begin{threeparttable}
		\caption{Balance between Treatment and Control groups at Baseline} \label{t:balanceall}
		\footnotesize
		\input{content/ttest_v3.tex}
		\begin{tablenotes}[flushleft]
			\item \emph{Notes:} Data from Baseline household survey. Unit of observation: household. Panel A: sample includes only households also surveyed at endline. Panel B: sample includes all households surveyed at baseline. Improved toilets refer to toilets of the quality defined using the classification in \cite{unicef2015}. For a detailed description of household and community-level covariates, please see Appendix \ref{Apsec:vardesc}.%\textit{*p \textless 0.10, ** p \textless 0.05, *** p \textless 0.01}.
		\end{tablenotes}
	\end{threeparttable}
\end{table}

The first panel of Table \ref{t:balanceall} reports household characteristics. Household heads in study communities are predominantly male (64\%), with a mean age of 55 years. Most completed primary school (68\%) and were employed (77\%). Average household size was four, with 30\% living with a child under the age of six. For 45\%, farming was the main activity. We also present a household asset wealth indicator, measured as the first factor of a principal component analysis based on a series of questions regarding asset ownership.\footnote{ Details of its components and their factor loadings are provided in Table \ref{t:rwi} in Appendix \ref{Apsubsec:hhchars}.} While the number value of this variable is in itself not meaningful, it is balanced across experimental arms.

The second panel presents the main outcomes, all measured at the household level. Our primary interest lies in households' sanitation behavior, in particular open defecation practices, for which we use two measures. The first captures whether the main survey respondent (typically a woman) states that at least one household member above the age of four years performs OD. The second variable takes the value one if the respondent herself declares to perform OD.\footnote{ Both rely on the question ``Where do you go to defecate?'', combined with a showcard of possible places. Urination habits are asked separately at baseline, but are not used here.} Table \ref{t:balanceall} shows that in our study population, 62.8\% of households have at least one member defecating in the open and a very similar percentage (62.4\%) of women report that they follow this practice themselves. 

We find that the habit of defecating in the open is closely linked to lack of toilet ownership. Ownership and use of private toilets are the most frequently discussed channels to reduce OD in CLTS community meetings \citep{kar2003}. Yet, only 36.9\% of households own a toilet, 36.1\% own a functioning toilet of any type, and 32.4\% own a functioning and improved toilet at baseline. We include all three measures in our analysis as outcomes since they capture different dimensions of interest. The first records ownership, but ignores functionality. The second additionally captures maintenance investments into the existing stock of toilets, or toilet divestment through lack of maintenance. The third accounts for quality, satisfying the stricter criteria set by the WHO/UNICEF Joint Monitoring Program regarding improved sanitation.

To validate households' reports of toilet ownership, interviewers asked respondents whether they may inspect the reported latrines at the endline interviews. If systematic measurement error was present in the treatment group as a result of CLTS, particularly due to over-reporting of ``desirable'' outcomes, then we would expect lower consent rates and higher rates of corrections in treatment relative to control areas at endline. However, we find this not to be the case: 24\% of households withheld their inspection consent, with no difference between treatment and control groups (24.86\% in control; 23.85\% in the treatment group, p-value=0.693). Inspection did not yield significant discrepancies between reported and actual ownership, and, if anything the correction rate is lower in the treatment group (7.98\%) than in control (9.56\%). None of these measures are statistically significantly different between treatment arms. We conclude that there is no evidence of selective reporting differences between treatment and control group in either dimension  - consent rates for inspection and truthful reports of ownership by consenting households. In additional sensitivity analysis, we exclude households who refused the validation of their toilets, and re-estimate average and heterogeneous impact estimates; all parameter estimates are virtually unchanged in the restricted sample.

CLTS impacts on reported open defecation practice post intervention, which the interviewers cannot monitor directly, are closely mirrored by reverse changes in (partially verifiable) toilet ownership (see results in Section 4.2).  This is in line with the close relationship observed between ownership and OD behaviour at baseline: 98\% of (functioning) toilet owners report that their household does not practice OD (see Table \ref{t:balanceall}). This percentage remains very similar after the intervention (95\%), and does not differ between households in treatment and control groups (p-value=0.345).\footnote{ This is different from OD habits in India for example, where toilet ownership and usage does not necessarily go hand in hand (see for example \citet{GuptaEtAl2019})}

The third panel of Table \ref{t:balanceall} finally presents statistics for a set of community characteristics. The communities are located on average about seven kilometers from the LGA head quarters (HQ)\footnote{ This is a walk of about an hour.} and encompass 1,616 households within a five kilometer radius. Part of our analysis, described in detail in Section \ref{sec:het}, will focus on heterogeneity in CLTS effectiveness by community socio-economic status (SES). Both the distance to the LGA HQ and number of households within a 5k radius are often used as proxies for SES status: distance to urban or semi-urban centers is typically used as a measure of remoteness or isolation, and the number of households within a 5km radius, as a measure of population density. We present here, and use later, summary statistics of two further SES proxies, namely a community wealth index based on the median household wealth in the community,\footnote{ Detailed lists of household asset items are frequently elicited in household surveys in developing countries, as they are often more precise than measures of household income. The aggregated index is mean zero and has a standard deviation of one.} and a pre-intervention nightlight intensity index within a 5km radius, a proxy for local economic wealth and income.\footnote{ \citet{Michalopoulos2013} presents evidence that wealth and nightlight intensity are strongly correlated.} We find that average nightlight intensity in our study area is very low with a mean of two relative to the total nightlight range of 0 to 61.

The next set of community characteristics we present relate to social interactions within the community, a dimension that has also been suggested as accelerating the effectiveness of CLTS (see for example \citet{Cameron2019}). We capture social interactions in three ways: a community's level of i) trust, ii) social capital, and iii) religious fragmentation. Trust is the average community score of household measures of the degree to which they trust their neighbors. Social capital is constructed similarly, based on households' participation in community events. We adapt the measures used in studies of ethnolinguistic fragmentation (ELF) to capture religious fragmentation, as our study sample is homogeneous along ethnic lines but very diverse in terms of religion. Detailed definitions of these measures and their distribution can be found in Appendix \ref{Apsubsec:commchars}. We find that religious fragmentation at 60\% was higher than in 70\% of the countries recorded in Alesina's 2003 fractionalization dataset, and people somewhat trusted their neighbors (i.e about 0.9 on a 0 to 2 scale).\footnote{ Mean social capital and community wealth are by construction close to zero.} 

Finally, in line with household level averages, the mean toilet ownership rates in our study clusters is 36.2\%. All community characteristics are balanced across treatment and control.

%% file: content/ttest_v3.tex
{
\def\sym#1{\ifmmode^{#1}\else\(^{#1}\)\fi}
\begin{tabular}{l*{1}{cccccc}}
\toprule
    &  All & \multicolumn{3}{c}{Control} & \multicolumn{2}{c}{Treatment-Control} \\
    \cmidrule(lr){3-5} \cmidrule(lr){6-7}
    & Obs. & Obs. & Mean & SD & Coeff. & \textit{p}-value \\
\midrule
\emph{Panel A - Postattrition household sample} & & & \\
\addlinespace
\emph{Household Characteristics} & & & \\
\addlinespace
HH head male (\%)& 4,014    & 2,027   & 64.53   & 47.85   & -1.92   & 0.307 \\
\addlinespace
HH head age (years)& 4,014    & 2,027   & 55.82   & 17.23   & -0.66   & 0.364 \\
\addlinespace
HH head employed (\%)& 4,014    & 2,027   & 78.10   & 41.37   & -0.79   & 0.677 \\
\addlinespace
HH head finished primary school (\%)& 4,014    & 2,027   & 67.54   & 46.83   &  0.40   & 0.851 \\
\addlinespace
Household size & 4,014    & 2,027   &  4.33   &  2.50   & -0.27   & 0.022 \\
\addlinespace
Household has at least 1 child below 6 y/o (\%)& 4,014    & 2,027   & 30.64   & 46.11   & -0.49   & 0.792 \\
\addlinespace
HH primary activity is farming (\%)& 4,014    & 2,027   & 47.11   & 49.93   &  3.06   & 0.414 \\
\addlinespace
Asset wealth index score &  4,014    & 2,027   &  0.05   &  2.04   & -0.02   & 0.879 \\
\addlinespace
\addlinespace
\emph{Open Defecation and Toilet Ownership}&            &            &                     \\
\addlinespace
At least 1 member (> 4 y/o) performs OD (\%)& 4,014    & 2,027   & 62.80   & 48.35   &  0.66   & 0.838 \\
\addlinespace
Main respondent performs OD (\%)& 3,974    & 2,008   & 62.40   & 48.45   &  0.47   & 0.884 \\
\addlinespace
Own a toilet (any condition, any type) (\%)& 4,014    & 2,027   & 36.90   & 48.27   & -0.31   & 0.922 \\
\addlinespace
Own a functioning toilet (any type) (\%)& 4,014    & 2,027   & 36.11   & 48.04   & -0.18   & 0.955 \\
\addlinespace
Own a functioning, improved toilet (\%)& 4,014    & 2,027   & 32.36   & 46.80   &  0.40   & 0.896 \\
\addlinespace
All members of the HH use a toilet (\%)& 4,014    & 2,027   & 37.20   & 48.35   & -0.66   & 0.838 \\
\addlinespace
All members of the HH use a toilet (cond. on ownership) (\%)&  1,446    &   732   & 97.95   & 14.18   & -0.89   & 0.35  \\
\addlinespace
\addlinespace
\emph{Community Characteristics}&            &            &                     \\
\addlinespace
Distance to the nearest LGA HQ, in km &  245    &   121   &  6.60   &  4.14   &  0.56   & 0.295 \\
\addlinespace
Number of households, 5km radius &  245    &   121   & 1,590  & 997  &-90  & 0.476 \\
\addlinespace
Community wealth &  245    &   121   & -0.21   &  0.95   & -0.13   & 0.353 \\
\addlinespace
Night light intensity, 5km radius, 2013 (min = 0, max = 25) &   245    &   121   &  2.07   &  3.01   & -0.06   & 0.870  \\
\addlinespace
Mean trust in neighbours (0-None, 2-High) &  245    &   121   &  0.88   &  0.49   & 0.01   & 0.847 \\
\addlinespace
Social capital & 245    &   121   & -0.09   &  1.10   &  0.16   & 0.252 \\
\addlinespace
Religious fragmentation (0-Low, 1-High) & 245    &   121   &  0.61   &  0.16   &  0.02   & 0.465 \\
\addlinespace
Mean toilet ownership rate (\%) & 245    &   121   & 36.24   & 24.51   & -1.32   & 0.674 \\
\addlinespace
\addlinespace
\emph{Panel B - Attrition}&            &            &                     \\
Not surveyed at endline (\%) & 4,540 & 2,281 & 11.14   & 31.46   &  0.91   & 0.341  \\
\bottomrule
\end{tabular}
}

%% file: 5.IEresults_new.tex
\section{Estimation approach}\label{sec:estimation}

We estimate the impacts of CLTS on our primary outcome, open defecation practices, using an intent-to-treat (ITT) design based on cluster randomized assignment to treatment.\footnote{ Intent-to-treat designs are informative about the key parameters of interest and often carry more external validity than estimates based on treatment recipiency, since perfect take-up of interventions is rare. As \citet{abramovsky2016} discuss, while leaders in all communities were approached, triggering meetings were not held in 18 communities (14\% of all treatment communities) due to insufficient number of community members coming to the planned event. In a successfully randomized scenario, as is our study (see Table \ref{t:balanceall}), ITT designs yield unbiased estimates of the average impact of the intervention on the sample assigned to treatment. For robustness purposes, in Appendix \ref{Apsubsec:attrition} we show that there is no evidence of selective triggering in our study. Additionally, we follow \cite{Imbens1994IdentificationEffects} and \cite{Angrist1995Two-stageIntensity} and instrument triggered treatment with treatment assignment. The results are very similar to the ITT estimates (see Appendix \ref{Apsubsec:attrition}).}  
We compare open defecation practices $y_{ict}$ in household $i$ living in community (cluster) $c$ in period $t$ by treatment assignment:
\begin{equation} \label{eq:Spec2}
    y_{ict} = \alpha + \gamma T_{c} + X_{ic0} \beta  + \theta y_{ic0} + \omega_{g} + \delta_{t} + \epsilon_{ict} 
\end{equation}
where community-level CLTS treatment status is defined by $T_{c}$. Baseline characteristics of households and their heads, $X_{ic0}$, are included to maximize precision and account for the imbalance in household size observed at baseline. To filter out unobserved area effects and contemporaneous shocks, we include LGA and survey wave fixed effects, $\omega_{g}$ and $\delta_{t}$. The parameter of interest, $\gamma$, captures the average impact of CLTS.

Our preferred specification conditions on the baseline value of the outcome variable, $y_{ic0}$. The resulting ANCOVA estimates are more efficient than difference-in-difference and simple difference estimators in experimental contexts,  when pre-treatment information is available and the outcome is strongly correlated over time (\cite{mckenzie2012beyond}). Alongside, we present conventional difference-in difference estimates.

To investigate heterogeneous impacts, we expand the specification in Equation (\ref{eq:Spec2}):
\begin{equation} \label{eq:Spec4}
    y_{ict} = \alpha + \gamma_{r} T_{c} + \gamma_{d} (T_{c} \times CC_{c}) + \phi CC_{c} + X_{ic0} \beta  + \theta y_{ic0} + \omega_{g} + \delta_{t} + \epsilon_{ict} 
\end{equation}
We introduce a binary variable $CC_{c}$ indicating a community characteristic, say community wealth, split our sample of communities along its median, and include the interaction term $T_{c} \times CC_{c}$. $\gamma_{r}$ is the average CLTS treatment effect in the richer half of communities (i.e. those for which $CC_{c}=0$), and $\gamma_{d}$ is the difference in treatment effects between rich and poor communities (i.e. those for which $CC_{c}=1$). The communities in our sample are typically located towards the middle (4th to 7th decile) of the Nigerian community wealth distribution (see Appendix \ref{Apsec:dhscomp}), rather than in the tails. Hence, neither communities are rich nor very poor relative to the Nigerian distribution. We simply denote the upper (lower) half of our study communities in the remainder of the text as ``rich'' (poor) communities for ease of description,  

Since we are testing multiple hypotheses simultaneously in our analysis of heterogeneous impacts, we report \emph{p}-values that are adjusted for the family-wise error rate in brackets. We compute these using the methodology proposed by \cite{romano2005stepwise}. %and the same cluster-bootstrapping procedure
Naive or unadjusted \emph{p}-values obtained from individual significant tests for each point estimate, calculated by drawing 1,000 clustered bootstrapped samples, are shown in parentheses.

\section{Results}\label{sec:results}
\subsection{Average impacts}
%NOTE (FO): The tables in this section include both naive (unadjusted) and robust (MHT adjusted) p-values. They do not include significance stars to let the reader make their own conclusions (and this is now the policy at AEA journals, for example: https://www.aeaweb.org/journals/aer/submissions/accepted-articles/styleguide). As a side-note, I'm using p-values in the Tables instead of standard errors although the AEA style guide linked above suggests including point estimates and se's only. But when we calculate the romano wolf adjusted significance, we can only obtain p-values. MHT adjusted standard errors can not be calculated by any code I (or anybody I've talked to) know of. I'm not even sure MHT adjusted standard errors make conceptual sense, given that the adjusted p-values are estimated from a null distribution of point estimates. So as a second best approach I think including both naive and adjusted p-values is good, because it helps to see the comparison between the two. We can always have a conversation with editors about this, and if necessary, include the naive (but clustered) standard errors as well into the tables, between the point estimates and the p-values, which I haven't done here because I think its too much info.  --- BRITTA: If someone like Orazio referees, then we will be asked to put se's. We could consider having both p-values and se's. But yes, happy to leave this for a later discussion.

We show average impact estimates on OD practices in Table \ref{t:baseline}. The first dependent variable is a dummy equal to 1 if at least one household member above the age of four performs OD (columns 1-3)\footnote{ Note that this question was not asked at the first follow-up survey, denoted FU1.}, the second a dummy variable equal to 1 if the main respondent performs OD (columns 4-6). Columns 1-2 (4-5) present difference-in-difference estimates without (with) household characteristics. Columns 3 and 6 present the ANCOVA specification described in equation \ref{eq:Spec2}, including household characteristics. 

% Table of basic results - pooled & period
\begin{table}[ht]
	\centering
	\begin{threeparttable}
		\caption{CLTS impacts on open defecation} \label{t:baseline}
		\footnotesize
		\input{content/table_base_lock.tex}
		\begin{tablenotes}[flushleft]
			\item \emph{Notes:} OD prevalence for any household member (columns 1-3) includes only members above the age of four, and was not measured in the first followup survey wave. \textbf{Household controls:} age, age squared, gender, education attainment level and employment status of the household head; household size, wealth asset score, and a dummy variable equal to one if farming is the household's main economic activity. Standard errors are clustered at the community level. \emph{p}-values are shown in parenthesis.
		\end{tablenotes}
	\end{threeparttable}
\end{table}

Panel A pools observations across the three followup surveys. We find that CLTS reduced OD consistently across all specifications and for both OD measures. Yet, the magnitude of behavioural change is small (around 3pp). As expected, the estimated coefficients are identical across specifications, but precision is highest when accounting for the lagged dependent variable. This being our preferred specification, we reject the null hypothesis of zero impact at the 10\% level for both measures of OD: exposure to CLTS resulted in a reduction in OD by 3pp. 

%(see equation \ref{eq:Spec3}). 
Due to possible divestment in toilet maintenance, as discussed in section \ref{sec:outcomes}, or short-lived behavioral changes in OD, CLTS impacts may not be sustained over time. We thus investigate time dynamics in impacts, considering impacts estimates approximately 8 (FU1), 24 (FU2), and 32 (FU3) months after CLTS implementation.  Panel B of Table \ref{t:baseline} reveals that point estimates are of similar magnitude throughout the period of study, and are statistically significant for the first and second followup waves. These estimates point towards a short-run reduction in OD eight months after CLTS. This reduction is sustained over two years after intervention implementation, but then fades out.

%%%%%%%%%%%%%%%%%%%%%%%%%%%%%%%%%%%%%%%%%%%%%%%%%%%%%%%%%%%%%%%%%%%%%%%%%%%%%%%%%%%%%%%%%%%%%%%%%%%%%
%%% HETEROGENEOUS IMPACTS on OD
%%%%%%%%%%%%%%%%%%%%%%%%%%%%%%%%%%%%%%%%%%%%%%%%%%%%%%%%%%%%%%%%%%%%%%%%%%%%%%%%%%%%%%%%%%%%%%%%%%%%%

\subsection{Heterogeneous impacts across communities}\label{sec:het}

%\subsubsection{Wealth and other measures of socio-economic status}
CLTS is designed and implemented as a participatory intervention at the community-level, with the aim of bringing about collective change. This raises the question of whether community characteristics hinder or foster intervention effectiveness. In spite of its current popularity, there is still scant experimental evidence where and under which conditions CLTS works. If CLTS is more (or only) effective in certain settings, successful targeting requires an understanding of the characteristics that best predict its effectiveness. 

The CLTS Handbook, a practitioners' guide drawing on field experience from 16 countries, suggests that the impact of CLTS on sanitation outcomes may depend on the socio-economic status (SES) of treated communities \citep{kar2008handbook}. It posits that successful implementation of CLTS is more likely in rural communities that are small, culturally and socially homogeneous, are located in remote areas and have a high prevalence of OD. In addition, \citet{Cameron2019} emphasize communities' social capital as a key facilitator for CLTS impact. Studies in the field of experimental economics have drawn similar conclusions, highlighting the importance of community rather than individual level differences in determinants of variations in concepts such as fairness and altruism \citep{HeinrichEtAl2006,HeinrichEtAl2001}, a conclusion made also in the context of Nigeria \citep{GowdyEtAl2003}.

Following these hypotheses, we define four broad indicators of local socio-economic status that may mediate CLTS impacts. In Section \ref{sec:outcomes}, we introduced community wealth as a widely available, comprehensive proxy for local socio-economic status (SES). We explore whether CLTS impacts vary by this SES proxy as well as using three additional proxy measures:  i) night light intensity, ii) population density, and iii) isolation.\footnote{ Details on all four measures are available in Appendix \ref{Apsubsec:commchars}.}

The pairwise correlations between the four characteristics demonstrate that poor communities, i.e. those below median wealth, are indeed often remote, less densely populated and have lower night light activity (see Table \ref{t:commcorr}). However, some pairwise correlations are relatively low, suggesting that each measure may capture a different aspect of socio-economic conditions. For instance, while population density and isolation appear to be highly correlated (rho = -0.58), the correlation between community asset wealth and average night light intensity is lower (rho = 0.16).\footnote{ Note that the inverse relationship between isolation and density reflects that higher population density is associated with shorter distances to the nearest LGA capital.} 

% Pairwise correlations between our SES proxy variables at the TU level (discrete)
\begin{table}[ht] 
    \centering
    \begin{threeparttable}
    \caption{Pairwise correlations of community level SES measures} \label{t:commcorr}
    \footnotesize
    \begin{tabular}{ l  c  c  c  c  c }
    \hline
    \textbf{SES measures}  & Asset wealth & Night lights & Density & Isolation & Toilet coverage \\ 
    \hline 
    Asset wealth & 1 & - & - & - & - \\
    Night lights & 0.1579 & 1 & - & - & - \\
    Density & 0.2524 & 0.4519 & 1 & - & - \\
    Isolation & -0.3434 & -0.4871 & -0.5762 & 1 & -\\
    Toilet coverage & 0.5459 & 0.2211 & 0.3523 & -0.3273 & 1 \\
    \hline
    Number of communities & \multicolumn{5}{c}{247} \\
    \hline
    \end{tabular}
    \begin{tablenotes}[flushleft]
			\item \emph{Notes:} Pairwise correlations of our four alternative measures for community SES. Details on how we construct these measures can be found in Appendix \ref{Apsec:vardesc}.
		\end{tablenotes}
    \end{threeparttable}
\end{table}

Additionally, we test whether the adjustment margin is higher in communities with low baseline toilet coverage, rendering them more susceptible to CLTS. %second, relative utility concerns may prompt higher sanitation investments in these areas. 
As expected, Table \ref{t:commcorr} shows that higher levels of SES measures are correlated with greater toilet coverage at the community level.

Table \ref{t:tuhte} presents the heterogeneous impact estimates, expressed as percentage point changes in OD, calculated using the pooled sample. The outcome variable captures the main respondent's OD practice, as this outcome is measured in all survey waves. Each column presents heterogeneous impacts by one of the four community-level SES measures and baseline toilet coverage, which we discretize along the sample median. For example, in column 1 we rank communities according to their wealth score. Communities with wealth scores equal to or above the median are defined as `High asset wealth' communities ($CC_c=0$), while the rest are classified as `Low asset wealth' communities ($CC_c=1$). The table shows the regression estimates for $\gamma_{r}$ and $\gamma_{d}$ from specification \ref{eq:Spec4}, as well as for the linear combination of both, for comparison purposes.
 
\begin{table}[ht]
	\centering
	\begin{threeparttable}
		\caption{Community economic conditions and CLTS impacts on OD} \label{t:tuhte}
		\footnotesize
		\input{content/tableTUHTE_SES_lock.tex}
		\begin{tablenotes}[flushleft]
			\item \emph{Notes:} All specifications control for the household and household head characteristics listed in Table \ref{t:baseline}. Errors are clustered at the community level. Naive (unadjusted) p-values shown in parenthesis. In brackets we present p-values adjusted by family wise error rate following \cite{romano2005stepwise}, using 1,000 cluster bootstrap samples and estimated jointly for all regressions presented in Tables \ref{t:tuhte}, \ref{t:tuhte3} and \ref{t:tuhte4}.
		\end{tablenotes}
	\end{threeparttable}
\end{table}

Table \ref{t:tuhte} shows strikingly consistent heterogeneous CLTS impacts for our four proxy measures of community SES. CLTS reduced OD prevalence by 7-9pp in communities with low asset wealth, low night light intensity, low density of households, or in communities that are far away from administrative capitals (i.e. high isolation, note the scale reversal here). Impact point estimates in these communities are two to three times as large as those found on average (Table \ref{t:baseline}) and remain statistically significant at endline. In contrast, we find statistically insignificant impact estimates close to zero in richer communities, regardless of whether we proxy these by asset wealth, night light intensity, density or isolation (see first row, columns 1-3 and second row, column 4). 

The third row of Table \ref{t:tuhte} presents the point estimates for the difference in CLTS impacts between the two halves of the sample, i.e. the estimated parameter $\gamma_{d}$ in Equation (\ref{eq:Spec4}). We reject the hypothesis that CLTS had the same impact on communities below and above the median for wealth, and find that the difference in CLTS impacts between poor and rich communities was 10pp (see column 1). Similarly, we reject it for all SES measures according to naive p-values at the 5\% level or lower. Using multiple hypothesis testing, we reject it for wealth and, marginally, for density. These results suggest that CLTS was indeed effective in a sub-sample of communities that shared underlying characteristics related to low SES. 

Asset wealth is highly correlated with toilet coverage at baseline. One possible explanation for our findings in columns 1-4 of Table \ref{t:tuhte} could therefore be that we are picking up differences in initial toilet coverage, and that the latter is the more relevant dimension of CLTS heterogeneity. This does not seem to be the case. In column 5 we find that the impact of CLTS on OD is 4pp stronger in areas with low initial toilet coverage areas. This is smaller than what we found using any of the SES proxies and is not significantly different from zero. Hence, the four SES proxies, in particular asset wealth, are a more informative measure to understand CLTS effectiveness than toilet coverage.

\medskip
The significant impact on OD behavior in poor communities might be driven by either a direct change in sanitation behavior, i.e. toilet usage, or by an increase in sanitation investments. Using the detailed measures of toilet stock, flow and quality, described in section \ref{sec:outcomes}, we break down sanitation investments into two components. First, CLTS may have promoted investment in new toilets, increasing toilet coverage (and its usage). Second, households who owned toilets at baseline may have invested more heavily into their maintenance and upkeep. This would increase the stock of \emph{functioning} toilets, reducing the depreciation of the stock of toilets. 

CLTS may also have changed sanitation behavior directly, either through increased toilet usage conditional on ownership, or through an increase in shared usage, i.e. when the main respondent declares to use a shared toilet (e.g. owned by a neighbor, public toilet or toilet at school or work).   

% Channels of OD reduction
\begin{table}[ht]
	\centering
	\begin{threeparttable}
		\caption{CLTS impacts on toilet ownership and usage} \label{t:margins}
		\footnotesize
		\input{content/table_channels_lock.tex}
		\begin{tablenotes}[flushleft]
			\item \emph{Notes:} All specifications control for the household and household head characteristics listed in Table \ref{t:baseline}. \emph{p}-values are shown in parenthesis. Standard errors are clustered at the community level and are adjusted for family-wise error rate following \cite{romano2005stepwise}, using 1,000 cluster bootstrap samples. 
		\end{tablenotes}
	\end{threeparttable}
\end{table}

% Impacts by toilet ownership at BL
% \begin{table}[ht]
% 	\centering
% 	\begin{threeparttable}
% 		\caption{CLTS impacts by baseline toilet ownership} \label{t:margins}
% 		\footnotesize
% 		\input{content/table_margin.tex}
% 		\begin{tablenotes}[flushleft]
% 			\item \emph{Notes:} All specifications control for the household and household head characteristics listed in Table \ref{t:baseline}. \emph{p}-values are shown in parenthesis. Standard errors are clustered at the community level and are adjusted for family-wise error rate following \cite{romano2005stepwise}, using 1,000 cluster bootstrap samples. 
% 		\end{tablenotes}
% 	\end{threeparttable}
% \end{table}

Table \ref{t:margins} presents impact estimates on sanitation investment and shared usage, showing pooled results in Panel A and heterogeneous impact estimates in Panel B. The reduction by 9pp in OD observed in poor communities (i.e. communities with low asset wealth in Table \ref{t:tuhte}) is almost identically matched in column 1 by an increase in toilet ownership of 8pp, suggesting an increase in the stock of toilets. Ownership of functioning toilets, i.e. maintained stock, increased by 10pp. These results strongly suggest that the change in sanitation behavior is driven by increased sanitation investment, mainly due to an increase in the toilet stock (see column 2).\footnote{ To conduct further sensitivity analysis for potential measurement error in toilet ownership, we estimate a probit model of selective inspection consent by community and household characteristics. We find that if anything, measurement error would bias differences between poor and rich communities towards zero: richer households and households in richer communities are more likely to refuse consent, increasing the possibility of over-reporting of ownership in this group. Critically, there are no differences in refusal rates between treatment and control groups, even when split by wealth levels. In addition, we test the robustness of our estimates by restricting the analysis to households that give inspection consent, and find that our results above are virtually unchanged.} 

In contrast, usage of existing toilets (column 3) and shared usage (column 4) in poor communities increased by much less or not at all - the differential impact of CLTS by community wealth is not statistically significant. Appendix \ref{Apsubsec:channels} shows that these results are robust to using the three alternative measures of communities' socio-economic status. In sum, OD reductions brought about by CLTS were due almost exclusively to increases in toilet ownership. Usage of owned toilets was high, around 80\% (p-value 0.19), and not statistically different between the ones that had a toilet already at baseline and those that built a toilet after the baseline data collection.

% Dynamic impacts of CLTS on OD and toilet ownership
\begin{figure}
	\centering
	\caption{CLTS treatment effects on OD and toilet ownership over time} \label{f:dynamic}
	\includegraphics[width=\textwidth]{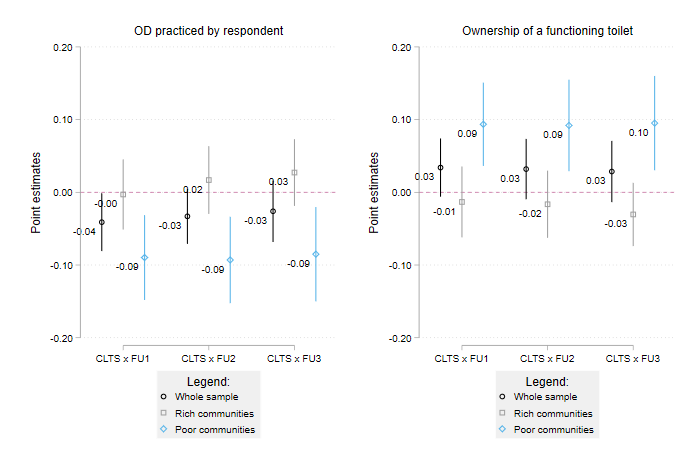}
	\caption*{\footnotesize{\emph{Note:} Graphs plot the point estimates for CLTS impacts by follow up survey wave and community level wealth. Results for the whole sample from the leftmost plot are equivalent to those presented in column 6 of Table \ref{t:baseline}. All specifications control for the household and household head characteristics listed in Table \ref{t:baseline}. Robust standard errors are clustered at the community level.}}
\end{figure}

Figure \ref{f:dynamic} shows dynamic CLTS impacts on toilet ownership and open defecation practice in poor and rich communities up to 32 months after the baseline survey. CLTS reduced OD in the short-run in poorer communities, and impacts were sustained over time (light blue). The estimated short- and long-run impacts of CLTS on both OD (see left panel) and toilet ownership (right panel) across the three followup periods are remarkably constant in poor communities. We draw two conclusions: First, our interpretation that CLTS-induced OD reductions are realized through increased toilet ownership holds also in the dynamic context. This suggests that CLTS has had a \textit{persistent} effect on OD in poor communities through toilet uptake, akin to that of a one-shot policy. Yet, given the strong link between toilet investment and improved sanitation behavior, CLTS impact in poor communities is achieved in the short-and sustained in the long-run.

\subsection{The role of community wealth: robustness}

In this section we conduct a number of robustness checks on our heterogeneous impact finding that wealth is a policy relevant margin for CLTS effectiveness.

\subsubsection{Functional form}

\begin{figure}[ht]
	\centering
	\caption{CLTS impacts by community wealth quartile} \label{f:tuwealthq}
	\includegraphics[width=.7\textwidth]{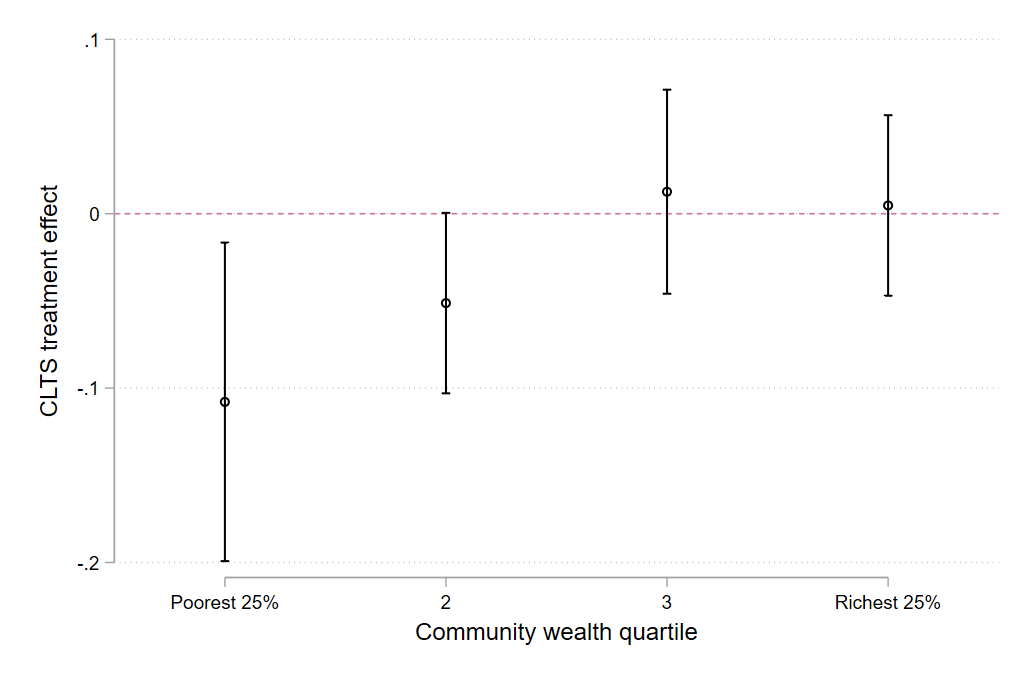}
	\caption*{\footnotesize{\emph{Note:} CLTS treatment effect coefficients by community wealth quartile. All specifications control for household characteristics listed in Table \ref{t:baseline}. Errors are clustered at the community level.}}
\end{figure}

We presented above differential impacts based on a discrete split into rich and poor communities. While using median values as cut-off is a standard approach%references?
, our results are qualitatively and quantitatively robust to alternative functional forms. Using for example a linear specification of community wealth rather than the discrete split into rich and poor communities, we find that treated communities that are one standard deviation poorer than the median display a 10\% reduction in OD (detailed results in Appendix \ref{Apsec:het}). Similarly, using quartiles of community asset wealth, we find that CLTS impacts are statistically significant and decreasing by wealth quartile (Figure \ref{f:tuwealthq}). They are statistically significant up to median wealth, for higher quartiles the treatment effects is zero. 

% CLTS impacts by TU wealth quartile

\subsubsection{Community versus household level heterogeneity}

Our measure of community asset wealth is an aggregation of household level information. Richer (poorer) communities tend to be composed of richer (poorer) households. Yet, significant heterogeneity between household and community wealth remains: 31\% of households living in poor clusters have higher asset wealth than the median, and 34\% of the households living in rich clusters are below median wealth. 

To understand whether our estimates are simply capturing that CLTS is (more) effective among poorer households (rather than poorer communities), we proceed in two ways. First, we run the same regression presented in column 1 of Table \ref{t:tuhte} but use a household wealth indicator instead of the community one. Second, we split the sample to investigate the effects of CLTS on OD for poor households in rich communities and rich households in poor communities. Appendix \ref{Apsubsec:commvshh} shows that CLTS is more effective among poorer households than rich ones, but the point estimate of the difference between rich and poor is about half as large as that of the community wealth impacts and not statistically significant under multiple hypothesis testing (see Column 1 in Appendix Table \ref{t:wealthhhvscomm}).\footnote{ In Table \ref{t:hhhte}, we also show that household composition or education do not explain the impact of CLTS.}  Furthermore, while in poor communities both rich and poor households reduce OD, there is no discernible effect nor a difference between poor and rich households in rich communities (see columns 2 and 3 in Appendix Table \ref{t:wealthhhvscomm}). This suggests that CLTS is more effective in poorer communities, regardless of the household's position in the wealth distribution.

\subsubsection{Characteristics of poor versus rich communities}

Poorer and richer communities may differ along many dimensions that may be relevant for the effectiveness of a program such as CLTS. In Table \ref{t:balanceTU} we present differences between the two groups at baseline by social interactions in the community, access to infrastructure and village leader characteristics.\footnote{ We in addition present in the Table what we refer to as `Household characteristics', the share of households headed by a male and with children below the age of 6 years. Since these variables do not differ between rich and poor communities, we do not discuss them further.}

\begin{table}
	\centering
	\begin{threeparttable}
		\caption{Baseline characteristics of rich and poor communities} \label{t:balanceTU}
		\footnotesize
		\input{content/ttest_RvP.tex}
		\begin{tablenotes}[flushleft]
			\item \emph{Notes:} All variables measured at baseline. Sample restricted to households interviewed at baseline and in all three followup survey waves. For a detailed description of household and community-level covariates, please see Appendix \ref{Apsec:vardesc}.  \textit{*p \textless 0.10, ** p \textless 0.05, *** p \textless 0.01}.
		\end{tablenotes}
	\end{threeparttable}
\end{table}

The index of social capital appears to be uncorrelated with community wealth. Poor communities, however, do exhibit higher levels of social cohesion (as measured by community-level trust and religious fragmentation) and lower levels of asset wealth inequality than rich communities.\footnote{ Details about how these measures are constructed are in appendix \ref{Apsubsec:commchars}.} We also find significant differences in access to public infrastructure, including having a local school, a hospital and paved internal roads. Finally, poor communities have less experienced and less educated leaders.

%At the same time, leaders play an important role in CLTS as they are the initial point of contact for the implementers and help organize the CLTS meeting in their village. Finally, they may also influence, through fines and other activities, whether collective action towards achieving defecation-free status is achieved. Yet, we find virtually no differences in CLTS effectiveness according to leader characteristics.  
We assess whether any of these dimensions might be the main drivers behind differential CLTS effectiveness by community wealth. \citet{kar2008handbook} suggest social interactions as a potential driver and \citet{Cameron2019} find that a CLTS intervention in Indonesia generated stronger effects in communities with higher social capital. However, results in Table \ref{t:tuhte3} show that the point estimates, in most cases, go in the opposite direction of what would be expected if stronger social cohesion improved program effectiveness. For example, we find slightly stronger reductions in OD in treated communities with lower social capital, high fragmentation and inequality. In all cases, differences in CLTS effectiveness along dimensions of social interactions are not significantly different from zero in our study context.

\begin{table}[ht]
	\centering
	\begin{threeparttable}
		\caption{CLTS impacts on OD by community level social interactions} \label{t:tuhte3}
		\footnotesize
		\input{content/tableTUHTE_OTHER_lock.tex}
		\begin{tablenotes}[flushleft]
			\item \emph{Notes:} All specifications control for the household and household head characteristics listed in Table \ref{t:baseline}. Errors are clustered at the community level. Naive (unadjusted) p-values shown in parenthesis. In brackets we present p-values adjusted by family wise error rate following \cite{romano2005stepwise}, using 1,000 cluster bootstrap samples and estimated jointly for all regressions presented in Tables \ref{t:tuhte}, \ref{t:tuhte3} and \ref{t:tuhte4}. 
		\end{tablenotes}
	\end{threeparttable}
\end{table}

We secondly explore whether our main results could be explained by poorer communities' lower access to infrastructure which may, for example, proxy for transport costs. Further, as leaders are the initial point of contact for the implementers and help organize the CLTS meeting in their village, leaders' tenure and education may play an important role. %\footnote{ They may also influence, through fines and other activities, whether collective action towards achieving defecation-free status is achieved.} 
We find no heterogeneous CLTS impacts along either of the three indicators of public infrastructure (Table \ref{t:tuhte4}, columns 1-3) nor along the village leader's tenure or education (Table \ref{t:tuhte4}, columns 4 and 5). \\

\begin{table}[ht]
	\centering
	\begin{threeparttable}
		\caption{CLTS impacts on OD by community infrastructure and village leader characteristics} \label{t:tuhte4}
		\footnotesize
		\input{content/tableTUHTE_OTHER2_lock.tex}
		\begin{tablenotes}[flushleft]
			\item \emph{Notes:} All specifications control for the household and household head characteristics listed in Table \ref{t:baseline}. Errors are clustered at the community level. Naive (unadjusted) p-values shown in parenthesis. In brackets we present p-values adjusted by family wise error rate following \cite{romano2005stepwise}, using 1,000 cluster bootstrap samples and estimated jointly for all regressions presented in Tables \ref{t:tuhte}, \ref{t:tuhte3} and \ref{t:tuhte4}. 
		\end{tablenotes}
	\end{threeparttable}
\end{table}

\subsubsection{Implementation heterogeneity}\label{sec:imphet}

CLTS is a fairly standardized intervention. Nonetheless, there is the possibility that the heterogeneous impacts we observe are the result of differences in the intervention's delivery. In recent work based on the same RCT, \cite{abramovsky2016} show that CLTS triggering meetings are more likely to fail, and not be carried out at all, in areas with high population density which is positively correlated with community wealth. Is CLTS ineffective in rich communities because CLTS triggering meetings are not taking place? Using community wealth directly, we find that the difference in triggering rates is small. The share of communities assigned to CLTS in which CLTS triggering meetings were successfully run was 75\% for rich communities, and 83\% in poor communities. These rates are not significantly different from each other at standard levels of statistical confidence (p-value=0.301). In addition, we present in Table \ref{t:implementation} Appendix \ref{Apsubsec:attrition} estimates of the impact of CLTS, instrumented by treatment assignment, and find very similar results to the ITT estimates.

Differences in delivery (and hence impacts) may also arise if meeting attendance rates are higher in poor communities. We do not find evidence for this hypothesis either. Attendance rates, measured as the number of attendees recorded by CLTS facilitators over village population, were not significantly different between rich and poor communities.\footnote{ On average, 34\% of the village members attended CLTS meetings in rich communities, while 42\% attended in poor communities. A community level regression of attendance rates on community level wealth group (i.e. a dummy equal to one if the community is poor) and LGA fixed effects results in a point estimate of just 3pp and a p-value of 0.662. Results are available upon request.} 

Differences in the quality of CLTS delivery may also arise from different delivery agents. As described in Section \ref{sec:design}, WaterAid hired two NGOs with CLTS experience, one from each state, to train local government officials in the facilitation of CLTS meetings. If these intermediate delivery agents (the NGOs) differed in the quality of their training activities, we might observe state-level differences in CLTS effectiveness.\footnote{ Of course, this is not the only channel that could explain differential results by state.} We find no evidence of this (see Table \ref{t:implementation} Appendix \ref{Apsubsec:attrition}). Interacting treatment status with state dummies in a specification similar to Equation (\ref{eq:Spec4}), we find that the interaction term is small and not significantly different from zero (p-value=0.326).

\subsection{Summary of findings} % not sure about this title but think it would be useful to separate this part a bit from the robustness section.

Taken together, our results suggest that community wealth encompasses a number of community characteristics that made CLTS more effective. We did not find evidence that any of these characteristics (such as toilet coverage, implementation or measures of social cohesion) could explain independently why CLTS worked better in poorer communities. 

While the available data does not allow us to pin down the drivers of these differential impacts, the heterogeneity dimension is in line with CLTS originators' theory and practitioners' experiences as to where the intervention should work best. The importance of context for CLTS effectiveness was also highlighted in a recent cross-country study, which concludes that ``[t]he impact of CLTS and subsequent sustained latrine use varied more by region than by intervention, indicating that context may be as or more important than the implementation approach in determining effectiveness'' \citep{CrockerEtAl2017Sust}.

In line with this observation, we will show in the next section that even without a context-specific understanding of underlying mechanisms, our heterogeneous CLTS impact can be used as a basis for more effective targeting of the intervention \textemdash within Nigeria and beyond. 
In other words, we will show that community wealth (or a proxy) is a precise predictor of CLTS effectiveness and can be used for policy targeting.

%In summary, we find strong evidence that CLTS, a community-level intervention, is effective in communities of lower socio-economic status, particularly in those with lower community wealth. In poor communities, OD rates decreased by 9pp from a baseline level of 75\% - a magnitude that is more than twice  as high as found in the average impact estimates. In contrast, we find no evidence of any impacts in richer communities.\\  

%[not quite sure where that goes - it goes against our findings for density]
%In a field experiment in Bangladesh, \citep{Guiteras2018} find that the social multiplier of individual subsidies for sanitation investment is larger in densely populated areas. 

%% file: content/table_base_lock.tex
{
\def\sym#1{\ifmmode^{#1}\else\(^{#1}\)\fi}
\begin{tabular}{l*{6}{c}}
\hline\hline
Outcome =1 if:      &\multicolumn{3}{c}{OD by any member}&\multicolumn{3}{c}{OD by the main respondent}                                        \\\cmidrule(lr){2-4}\cmidrule(lr){5-7}
					&\multicolumn{1}{c}{(1)}         &\multicolumn{1}{c}{(2)}         &\multicolumn{1}{c}{(3)}         &\multicolumn{1}{c}{(4)}         &\multicolumn{1}{c}{(5)}         &\multicolumn{1}{c}{(6)}         \\
\hline
\emph{Panel A - Pooled impacts}&                     &                     &                     &                     &                     &                     \\
[1em]
CLTS ($\gamma$)     &       -0.03         &       -0.03         &       -0.03  &       -0.03         &       -0.04         &       -0.03 \\
\hspace{5pt} \emph{p}-value &     (0.26)       &      (0.16)         &      (0.06)        &      (0.22)         &      (0.13)         &      (0.04)         \\
[1em]
\hline
\emph{Panel B - Dynamic impacts}&                     &                     &                     &                     &                     &                     \\
[1em]
CLTS x FU 1 ($\gamma$\textsubscript{1})&                 &               &                &       -0.04         &       -0.04         &       -0.04 \\
\hspace{5pt} \emph{p}-value                    &               &                &                &      (0.16)         &      (0.11)         &      (0.04)         \\
[1em]
CLTS x FU 2 ($\gamma$\textsubscript{2})&       -0.04         &       -0.04         &       -0.04  &       -0.03         &       -0.04         &       -0.03  \\
\hspace{5pt} \emph{p}-value                    &      (0.18)         &      (0.11)         &     (0.06)         &      (0.23)         &      (0.14)         &      (0.09)         \\
[1em]
CLTS x FU 3 ($\gamma$\textsubscript{3})&       -0.03         &       -0.03         &       -0.03         &       -0.02         &       -0.03         &       -0.03         \\
\hspace{5pt} \emph{p}-value                  &      (0.41)         &      (0.31)         &      (0.15)         &      (0.42)        &      (0.34)         &      (0.22)         \\
[1em]
\hline
ANCOVA               &        No         &        No         &        Yes         &        No         &        No         &        Yes         \\
Household controls   &        No         &        Yes         &        Yes         &        No         &        Yes         &        Yes         \\
DV control mean (EL)&        0.49&        0.49&        0.49&        0.48&        0.48&        0.48\\
No. of communities  &         246         &         246         &         246         &         246         &         246         &         246         \\
No. of observations &       8,786         &       8,518         &       8,518         &      13,233         &      12,830         &      12,697         \\
\hline\hline
\end{tabular}
}

%% file: content/tableTUHTE_SES_lock.tex
\begin{tabular}{l*{5}{c}}
\hline\hline
&\multicolumn{5}{c}{Dep. variable: main respondent performs OD}                 \\\cmidrule(lr){2-6}
Community Characteristic (CC) at BL:&\multicolumn{1}{c}{Asset wealth}&\multicolumn{1}{c}{Night lights}&\multicolumn{1}{c}{Density}&\multicolumn{1}{c}{Isolation}&\multicolumn{1}{c}{Toilet coverage}\\
                    &\multicolumn{1}{c}{(1)}&\multicolumn{1}{c}{(2)}&\multicolumn{1}{c}{(3)}&\multicolumn{1}{c}{(4)}&\multicolumn{1}{c}{(5)}\\
\hline
CLTS $\times$ High ($\gamma_{r}$)       &        0.01   &        0.00   &        0.01   &       -0.07   &       -0.01   \\
\hspace{5pt} \textit{p}-value (naive)   &      (0.50)   &      (0.86)   &      (0.73)   &      (0.01)   &      (0.50)   \\
\hspace{5pt} \emph{p}-value (MHT robust)&      [0.93]   &      [0.93]   &      [0.93]   &      [0.05]   &      [0.93]   \\
[1em]
CLTS $\times$ Low ($\gamma_{r} + \gamma_{d}$)   &       -0.09   &       -0.07   &       -0.08   &        0.00   &       -0.05   \\
\hspace{5pt} \textit{p}-value (naive)           &      (0.01)   &      (0.01)   &      (0.01)   &      (0.98)   &      (0.09)   \\
\hspace{5pt} \emph{p}-value (MHT robust)        &      [0.02]   &      [0.06]   &      [0.03]   &      [0.98]   &      (0.42)   \\
[1em]
Difference ($\gamma_{d}$)                       &       -0.10   &       -0.07   &       -0.09   &        0.07   &       -0.04   \\
\hspace{5pt} \textit{p}-value (naive)           &      (0.00)   &      (0.02)   &      (0.01)   &      (0.03)   &      (0.26)   \\
\hspace{5pt} \emph{p}-value (MHT robust)        &      [0.04]   &      [0.24]   &      [0.10]   &      [0.26]   &      (0.92)   \\
\hline
DV control mean (EL, High)  &        0.36   &        0.39   &        0.41   &        0.57   &        0.34   \\
DV control mean (EL, Low)   &        0.62   &        0.57   &        0.57   &        0.40   &        0.66   \\
No. of TUs          &         246   &         246   &         246   &         246   &         246   \\
No. of obs.         &      12,697   &      12,697   &      12,697   &      12,697   &      12,697   \\
\hline\hline
\end{tabular}

%% file: content/table_channels_lock.tex
{
\def\sym#1{\ifmmode^{#1}\else\(^{#1}\)\fi}
\begin{tabular}{l*{4}{c}}
\hline\hline
Outcome =1 if:      					&\multicolumn{1}{c}{Owns toilet}&\multicolumn{1}{c}{Owns functioning toilet}&\multicolumn{1}{c}{Usage (if functioning)}&\multicolumn{1}{c}{Shared use}\\
                    					&\multicolumn{1}{c}{(1)}         &\multicolumn{1}{c}{(2)}         &\multicolumn{1}{c}{(3)}         &\multicolumn{1}{c}{(4)}         \\
\hline
\emph{Panel A - Pooled impacts}			&                     &                     &                     &                    \\
[1em]
CLTS ($\gamma$)     					&        0.02         &        0.03         &        0.00         &        0.01         \\
\hspace{5pt} \emph{p}-value (naive) &      (0.25)         &      (0.07)         &      (0.83)         &      (0.25)         \\
\hspace{5pt} \emph{p}-value (MHT robust) &      [0.55]        &      [0.21]         &      [0.83]         &      [0.55]         \\
[1em]
\hline
\multicolumn{3}{l}{\emph{Panel B - By community wealth group}}                    &                     &                   \\
[1em]
CLTS x Rich ($\gamma_{r}$)              &       -0.03         &       -0.02         &       -0.01         &        0.01         \\
\hspace{5pt} \emph{p}-value (naive) &      (0.10)         &      (0.25)         &      (0.41)         &      (0.02)         \\
\hspace{5pt} \emph{p}-value (MHT robust)&      [0.23]        &      [0.42]         &      [0.42]         &      [0.07]         \\
[1em]
CLTS x Poor ($\gamma_{r} + \gamma_{d}$)	&        0.08         &        0.10         &        0.03         &       -0.00         \\
\hspace{5pt} \emph{p}-value (naive) &        0.01         &        0.01         &        0.16         &        0.99         \\
\hspace{5pt} \emph{p}-value (MHT robust) &      [0.01]        &      [0.01]         &      [0.31]         &      [0.99]         \\
[1em]
Difference ($\gamma_{d}$) 				&        0.11         &        0.12         &        0.05         &       -0.02         \\
\hspace{5pt} \emph{p}-value (naive) &      (0.00)         &      (0.00)         &      (0.11)         &      (0.37)         \\
\hspace{5pt} \emph{p}-value (MHT robust) &      [0.01]        &      [0.01]         &      [0.20]         &      [0.37]         \\
[1em]
\hline
DV control mean (EL, Rich)&        0.47&        0.46&        0.57&        0.04\\
DV control mean (EL, Poor)&        0.24&        0.24&        0.44&        0.02\\
No. of communities   					&         246         &         246         &         245         &         246         \\
No. of observations 					&      12,497         &      12,497         &       7,113         &      12,697         \\
\hline\hline
\end{tabular}
}

%% file: content/ttest_RvP.tex
{
\def\sym#1{\ifmmode^{#1}\else\(^{#1}\)\fi}
\begin{tabular}{l*{1}{ccc}}
\toprule
Community wealth group:&        Rich&        Poor&     P-value         \\
\midrule
&&& \\ \emph{Social interactions in the community}&            &            &                     \\
\addlinespace
\hspace{0.1cm} Social capital index (mean = 0, SD = 1)&       0.117&      0.0978&        0.88         \\
\addlinespace
\hspace{0.1cm} Trust in neighbours (0-None, 2-High, SD = 0.40)&       0.809&       0.970&        0.00\sym{***}\\
\addlinespace
\hspace{0.1cm} Religious fragmentation (0-Low, 1-High)&       0.642&       0.598&        0.03\sym{**} \\
\addlinespace
\hspace{0.1cm} Asset wealth inequality&       0.951&       0.612&        0.00\sym{***}\\
\addlinespace
&&& \\ \emph{Communities' public infrastructure}&            &            &                     \\
\addlinespace
%\hspace{0.1cm} OD prevalence (\%)&       53.41&       75.60&        0.00\sym{***}\\
%\addlinespace
%\hspace{0.1cm} Share of HHs with functioning toilets (\%)&       45.66&       23.58&        0.00\sym{***}\\
%\addlinespace
%\hspace{0.1cm} Improved drinking water available (\%)&       98.88&       99.15&        0.82         \\
%\addlinespace
\hspace{0.1cm} Has paved internal roads (\%)&       57.16&       28.91&        0.00\sym{***}\\
\addlinespace
\hspace{0.1cm} Has a local hospital (\%)&       23.24&       3.866&        0.00\sym{***}\\
\addlinespace
\hspace{0.1cm} Has a local primary school (\%)&       72.53&       61.18&        0.05\sym{**} \\
\addlinespace
%&&& \\ \emph{Community SES proxies}&            &            &                     \\
%\addlinespace
%\hspace{0.1cm} Night light intensity, 2013 (min = 0, max = 25)&       2.350&       1.697&        0.07\sym{*}  \\
%\addlinespace
%\hspace{0.1cm} Number of households within a 5km radius (SD = 989)&      1675.0&      1397.6&        0.03\sym{**} \\
%\addlinespace
%\hspace{0.1cm} Distance to the nearest LGA HQ, in km (SD = 4.13)&       5.712&       8.057&        0.00\sym{***}\\
%\addlinespace
&&& \\ \emph{Village leader characteristics}&            &            &                     \\
\addlinespace
\hspace{0.1cm} Years as leader&       11.59&       9.100&        0.04\sym{**} \\
\addlinespace
\hspace{0.1cm} Completed primary school (\%)&       60.42&       42.42&        0.00\sym{***}\\
\addlinespace
\hspace{0.1cm} Affiliated to a political party (\%)&       29.02&       33.20&        0.47         \\
\addlinespace
&&& \\ \emph{Household characteristics}&            &            &                     \\
\addlinespace
\hspace{0.1cm} Share of HHs with male heads (\%)&       65.25&       62.60&        0.19         \\
\addlinespace
%\hspace{0.1cm} Share of HHs heads with completed primary school (\%)&       75.11&       59.42&        0.00\sym{***}\\
%\addlinespace
\hspace{0.1cm} Share of HHs with children below 6 y/o (\%)&       29.38&       28.70&        0.71         \\
\midrule
Observations        &         233&            &            \\ \bottomrule
\end{tabular}
}

%% file: content/tableTUHTE_OTHER_lock.tex
\begin{tabular}{l*{4}{c}}
\hline\hline
&\multicolumn{4}{c}{Dep. variable: main respondent performs OD} \\\cmidrule(lr){2-5}
Community Characteristic (CC) at BL:                 &\multicolumn{1}{c}{Trust}&\multicolumn{1}{c}{Social capital}&\multicolumn{1}{c}{Fragmentation}&\multicolumn{1}{c}{Inequality}\\
                    &\multicolumn{1}{c}{(1)}&\multicolumn{1}{c}{(2)}&\multicolumn{1}{c}{(3)}&\multicolumn{1}{c}{(4)}\\
\hline
CLTS $\times$ High ($\gamma_{r}$)           &       -0.04   &       -0.01   &       -0.04   &       -0.05   \\
\hspace{5pt} \textit{p}-value (naive)       &      (0.11)   &      (0.49)   &      (0.08)   &      (0.10)   \\
\hspace{5pt} \emph{p}-value (MHT robust)    &      [0.53]   &      [0.93]   &      [0.47]   &      [0.53]   \\
[1em]
CLTS $\times$ Low ($\gamma_{r} + \gamma_{d}$)   &       -0.03   &       -0.05   &       -0.02   &       -0.02   \\
\hspace{5pt} \textit{p}-value (naive)                        &      (0.17)   &      (0.04)   &      (0.30)   &      (0.27)   \\
\hspace{5pt} \emph{p}-value (MHT robust)        &      [0.60]   &      [0.26]   &      [0.68]   &      [0.68]   \\
[1em]
Difference ($\gamma_{d}$)                       &        0.02   &       -0.04   &        0.02   &        0.03   \\
\hspace{5pt} \textit{p}-value (naive)                        &      (0.60)   &      (0.25)   &      (0.56)   &      (0.43)   \\
\hspace{5pt} \emph{p}-value (MHT robust)        &      [0.98]   &      [0.92]   &      [0.98]   &      [0.98]   \\
\hline
DV control mean (EL, High)  &        0.44   &        0.44   &        0.49   &        0.56   \\
DV control mean (EL, Low)   &        0.52   &        0.53   &        0.47   &        0.41   \\
No. of TUs          &         246   &         246   &         246   &         246   \\
No. of obs.         &      12,697   &      12,697   &      12,697   &      12,697   \\
\hline\hline
\end{tabular}

%% file: content/tableTUHTE_OTHER2_lock.tex
\begin{tabular}{l*{5}{c}} 
\hline\hline
&\multicolumn{5}{c}{Dep. variable: main respondent performs OD} \\\cmidrule(lr){2-6}
&\multicolumn{3}{c}{Public goods}               &\multicolumn{2}{c}{Leader} \\\cmidrule(lr){2-4}\cmidrule(lr){5-6}
Community Characteristic (CC) at BL:  &\multicolumn{1}{c}{Road}&\multicolumn{1}{c}{Hospital}&\multicolumn{1}{c}{School}&\multicolumn{1}{c}{Experience}&\multicolumn{1}{c}{Education}\\
&\multicolumn{1}{c}{(1)}&\multicolumn{1}{c}{(2)}&\multicolumn{1}{c}{(3)}&\multicolumn{1}{c}{(4)}&\multicolumn{1}{c}{(5)}\\
\hline
CLTS $\times$ Yes/High ($\gamma_{r}$)           &       -0.02   &       -0.04   &       -0.04   &       -0.04   &       -0.03   \\
\hspace{5pt} \textit{p}-value (naive)           &      (0.43)   &      (0.06)   &      (0.20)   &      (0.09)   &      (0.23)   \\
\hspace{5pt} \emph{p}-value (MHT robust)        &      [0.93]   &      [0.31]   &      [0.74]   &      [0.49]   &      [0.77]   \\
[1em]
CLTS $\times$ No/Low ($\gamma_{r} + \gamma_{d}$)&       -0.06   &       -0.01   &       -0.03   &       -0.03   &       -0.04 \\
\hspace{5pt} \textit{p}-value (naive)           &      (0.01)   &      (0.85)   &      (0.11)   &      (0.17)   &      (0.08)   \\
\hspace{5pt} \emph{p}-value (MHT robust)        &      [0.12]   &      [0.98]   &      [0.51]   &      [0.60]   &      [0.40]   \\
[1em]
Difference ($\gamma_{d}$)                       &      -0.04    &        0.03   &        0.01   &        0.00   &       -0.01   \\
\hspace{5pt} \textit{p}-value (naive)           &      (0.24)   &      (0.43)   &      (0.78)   &      (0.89)   &      (0.65)   \\
\hspace{5pt} \emph{p}-value (MHT robust)        &      [0.92]   &      [0.98]   &      [0.98]   &      [0.98]   &      [0.98]   \\
\hline
DV control mean (EL, High)  &        0.51   &        0.49   &        0.47   &        0.50   &        0.53   \\
DV control mean (EL, Low)   &        0.45   &        0.40   &        0.48   &        0.46   &        0.43   \\
No. of TUs          &         235   &         233   &         235   &         232   &         232   \\
No. of obs.         &      11,901   &      11,793   &      11,901   &      11,619   &      11,692   \\
\hline\hline
\end{tabular}

%% file: 6.comparisons.tex
\section{Intervention targeting and transferability}

It is widely advocated that decisions about investment in public health interventions should be evidence-based. Due to the high cost of producing location-specific evidence, policy decisions are usually based on a limited number of studies, often conducted elsewhere. Their results are generalized to make an implementation decision in a different set of target sites. This is despite the understanding that outcomes depend on both the types of interventions and the context in which they are implemented \citep{WangEtAl2006,Meager2018UnderstandingExperiments}. Or, as \citet{Angrist2004} put it `[t]he relevance or 'external validity' of a particular set of empirical results is always an open question.' Recent reviews of RCT impact evaluation studies, in the medical \citep{Malmivaara2019} as well as the development economics literature \citep{PetersEtAl2018}, highlight that their external validity may be limited.

At the same time, context-specific impacts may arise due to heterogeneous responses of population subgroups, leading to diverging average treatment impacts across RCTs that sample (randomly) from populations with different underlying characteristics. As such, a better understanding of heterogeneous responses could help strengthen the generalizability to and external validity of RCTs in different target populations \citep{HotzEtAl2005,AllcottMullainathan2012,ImaiRatkovic2013,FerraroMiranda2013}. Policy makers can use evidence on responsive subgroups for a more cost effective targeted implementation strategy \citep{HeckmanEtAl1997,DjebbariSmith2008}.

We argue, in the same spirit, that policy-makers in Nigeria and in other countries can build on our findings, target poor geographical areas and, by doing so, avoid wasting money on implementing CLTS in low or no response communities. This is particularly relevant given recent estimates that `the true costs of participatory sanitation' can be quite substantial \citep{CrockerEtAl2017}: existing cost estimates range from US\$30.34 to 81.56 per targeted household in Ghana, and US\$14.15 to 19.21 in Ethiopia. % BRITTA: I STILL DON'T FEEL THIS STREAMLINING ARGUMENT FOLLOWS NATURALLY. EITHER WE EXPLAIN MORE OR DELETE IS MY SUGGESTION also suggests that there may be scope for streamlining its implementation. 
We present our argument for the external validity of our findings to other contexts in the next section.

\subsection{Community wealth as a factor reconciling diverging CLTS RCT results}\label{sec:comparisons}

Developed in Bangladesh in 1999, CLTS is today widely implemented and endorsed by national governments and NGOs in more than 25 Latin American, Asian and African countries. Given limited evidence on its effectiveness in its early existence \citep{VenkataramananEtAl2018}, its spread was typically a reaction to enthusiastic advocacy by numerous actors, including grassroot activists, state bureaucrats and the donor community \citep{Zuin2019}.

Rigorous evaluation studies emerged in the early 2010s. Yet, even if policy-makers today wanted to base CLTS implementation decisions on the available evidence, the conflicting impact estimates would not be a useful guide. To the best of our knowledge, five other studies have estimated the impact of CLTS-like interventions in developing countries based on RCTs. The World Bank's Water and Sanitation Programme (WSP) conducted three of these studies, in Tanzania \citep{bricenoEtAl2017}, %BRITTA: Tanzania is the only country where we are not more specific about location. Should we just mention country here throughout? If not, I checked and Tanzania study takes place in 10 out of 129 districts. These are: Mpwapwa, Kondoa, Rufiji, Iringa, Sumbawanga, Kiteto, Masasi, Musoma, Karagwe and Igunga.
Madhya Pradesh, India \citep{Patil2014TheTrial}, and East Java, Indonesia \citep{Cameron2019}. \cite{pickering2015effect} conducted a similar evaluation in rural Mali. Finally, \cite{guiteras2015encouraging} carried out a cluster RCT in Tanore, Bangladesh, in which they evaluated the impact of three different policy approaches, one of which was a CLTS-style intervention.\footnote{ The other two were supply side technical assistance and subsidy provision.} The resulting evidence is inconclusive, with estimated CLTS impacts ranging from 30pp increases in toilet ownership in Mali to no statistically-detectable impacts in the studies conducted in Bangladesh and Indonesia. 

In this section, we argue that community-level SES, in particular wealth, might be the factor underlying these diverging results. In Section \ref{sec:results}, we documented the strong heterogeneity of CLTS impacts by community wealth in our field experiment in Nigeria. If these results are externally valid, CLTS interventions in richer areas or study sites will generally have lower (or no detectable) impact than those administered in poorer areas. 

Using study-specific data, we analyze how CLTS effectiveness across (and within) RCTs varied along SES status. As a first step, we use a consistent method to estimate heterogeneous impacts on toilet ownership and open defecation by community socio-economic status for each RCT where data is publicly available (Bangladesh, India, Indonesia and Tanzania).%BRITTA: SHOULD WE MENTION THAT IMPACT ESTIMATES ARE CONSISTENT WITH THOSE PRESENTED IN PUBLISHED STUDIES?
\footnote{The datasets for these studies can be found at: \cite{DVN/GJDUTV_2017} (Bangladesh), \cite{WaterandSanitationProgram2009WSP2009-2011} (India), \cite{WaterandSanitationProgram2008IndonesiaIDN_2009_2011_WSP-IE_v01_M_v01_A_PUF} (Indonesia), and \cite{Briceno2012ImpactTZA_2012_SHRSBIE-EL_v01_M_v01_A_PUF} (Tanzania). For the RCT in Mali (\citet{pickering2015effect}), where data is not publicly available, we report the estimates presented in their paper, obtained by a means comparison of outcomes between treatment and control households at endline. The OD outcome captures whether any adult female performed OD.}

  %While our preferred specification is ANCOVA with household level controls, other studies relied on simple-differences or first-differences estimates. 
%ML: In the next paragaph, we estimate homogeneous effects (whole sample) and heterogeneous impacts by splitting the samples into poor and richer half. This is somewhat confusing relative to the sentence above. So I changed this a bit:

Optimally, we would compare impact estimates along a community measure of wealth that is a) consistent across studies, and b) aggregated by treatment unit, i.e. community. Unfortunately, not all studies collected data on ownership of consumer durables that is used to construct the community wealth index in our Nigerian study. In addition, the durable items underlying asset wealth indices are highly country- and context-specific \citep{Filmer2001EstimatingIndia}.\footnote{Similar limitations rule out the use of population density and distance to the nearest district headquarters.} %BRITTA: I FEEL THIS NEXT SENTENCE IS A DIFFERENT POINT AND DOES NOT FIT HERE. SHALL WE DELETE?
Aggregation at treatment unit is not always possible due to data confidentiality reasons which made the exact location of study clusters non-disclosable in a majority of the cited studies. 

For these reasons, we use night light intensity at baseline as an alternative measure of community-level SES in estimation. Night light indices are comparable across locations, and we compute them at the lowest geographical level $d$ available in each study.\footnote{Appendix Table \ref{Apt:nightlev} gives a detailed overview of the level of geographical disaggregation and the number of observed geographical units used to calculate average night light intensities. In line with our analysis in the previous sections, we exploit the within-study variation in night light intensity to calculate treatment effects separately for the poorer and richer halves of each sample. See appendix \ref{Apsec:comp} for full details and exceptions where this procedure was not possible due to data limitations. All night light indices relate to the baseline survey year.} While used in the past to proxy for GDP \textit{per capita}, a measure of income, at the sub-national level in African countries \citep{Michalopoulos2013}, we have shown that it is a robust proxy for community SES, and yields very similar heterogeneous impact estimates to those obtained by community wealth (see Table \ref{t:tuhte}).\footnote{\cite{Michalopoulos2013} also presents evidence of a strong correlation between wealth and nightlight intensity.} 

For estimation, we split each RCT sample into richer and poorer clusters, and then implement a relatively parsimonious ANCOVA model based on information that is commonly available across studies, leading to the following specification:   

%OLD Based on information that is commonly available across studies we implement a relatively parsimonious ANCOVA specification. For each study, including our own, we separately estimate CLTS impacts for three samples - whole sample, households in richer clusters and households in poorer clusters - using the following regression model:
\begin{equation} \label{eq:compare}
    y_{icd} = \alpha + \gamma_{r} T_{cd} +  \theta \, y_{icd0} +  X_{icd} \, \beta + \omega_{d} + \epsilon_{icd} 
\end{equation}
The outcome variables, $y_{icd}$, are two dummy variables, denoting i) ownership of a private, functioning toilet by household $i$ in cluster $c$ in district $d$, and ii) OD by any member of the household. $T_{cd}$ is a dummy variable equal to one if cluster $c$ from district $d$ has been assigned to CLTS, and zero otherwise. We distinguish between district and cluster here, as for data protection reasons we observe treatment assignment at the anonymised cluster level $c$, and observe community SES at the non-anonymised district level $d$.\footnote{In cases where multiple treatment arms existed, we include only the CLTS arm and the control group. We include baseline values $y_{icd0}$ for the outcome variables, except for Tanzania, where no baseline data was collected. Our study in Nigeria is the only one to include multiple post-treatment survey waves, so we restricted our sample to observations from the second followup wave (FU2), which is closest in timing to the post-intervention surveys of the other studies.} Household level controls $X_{icd}$ are gender, age and age squared of the household head, and whether a household's primary activity is farming. We include geographic fixed effects at the district level ($\omega_{d}$) where appropriate, to account for stratified randomization.\footnote{ District level fixed effects are used to estimate impacts for the case of India, Indonesia, Nigeria and Tanzania. We omit geographical fixed effects in the case of Bangladesh since the experiment was conducted in a single district.} Standard errors are clustered at the cluster level, the unit of randomization. %The data from the RCT in Mali (\citet{pickering2015effect}) is not publicly available, so we use the estimates presented in their paper, obtained by a means comparison of outcomes between treatment and control households at endline. The OD outcome selected for this exercise is whether any adult female performed OD.

\begin{figure}
	\centering
	\caption{CLTS impacts on OD and toilet ownership by average night light intensity} \label{f:compare}
    \includegraphics[width=\textwidth] {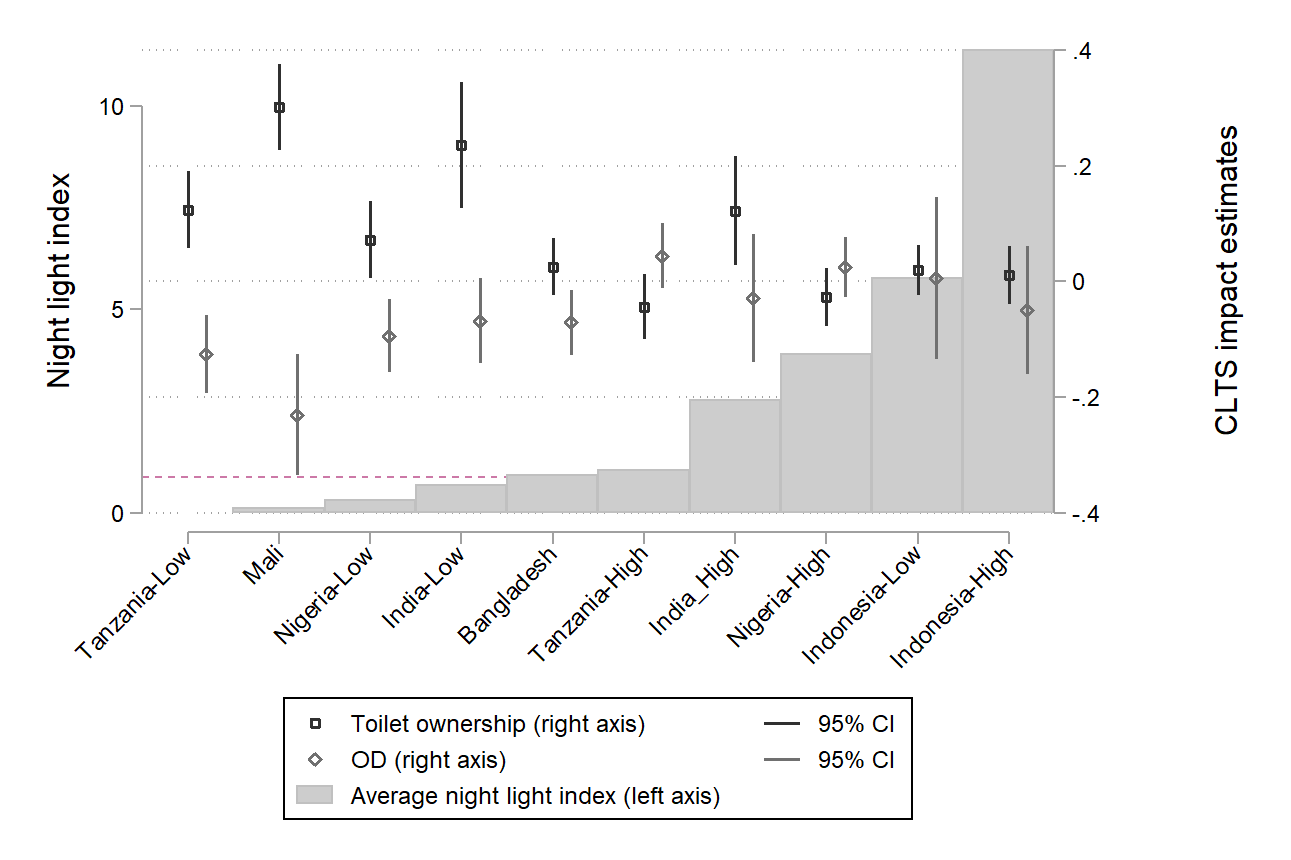}
     %{content/Comparison_onlyCountry.png}
	\caption*{\footnotesize{\emph{Note:} Study-specific point estimates from a simple differences regression of CLTS impacts on OD prevalence (gray diamonds) and toilet ownership (black dots). Gray bars show the average night light intensity recorded by NASA's Operational Line Scan (OLS) system, on the year of each study's baseline survey, over the study area. Data from the Mali study was unavailable, so results from Table 2 in \cite{pickering2015effect} were used instead (in the case of OD, results for adult women were used).}} 
\end{figure}

In Figure \ref{f:compare} we report the study-specific point estimates of CLTS impacts for poorer and richer communities on toilet ownership and open defecation (right axis) along with the corresponding satellite nightlight intensity over the considered set of study areas (left axis).\footnote{The exception is Mali where we cannot produce heterogeneous impact estimates due to the lack of data access.} It illustrates our main result, that CLTS is on average more effective in poorer settings, i.e. among those with lower nightlight intensity to the left of the Figure. As night light intensity increases, point estimates decrease in magnitude for both outcomes (depicted in black circles for toilet ownership and grey diamonds for OD), and the likelihood of rejecting the null hypothesis falls. Comparing the within-RCT impact estimates (e.g. for example Tanzania-Low versus High), we find that impact estimates are always larger in the `poorer' half of communities. Note that we find no evidence of CLTS impacts in study sites with night light intensity above that of the Nigerian median (i.e with average nightlight above 0.89, marked by the horizontal dotted line just above the 0 value of Night Light Index Axis). In contrast, all four community groups with the lowest nightlight indices (and all samples with average nightlight below the poorer Nigerian communities) display statistically significant CLTS impacts.  We argue that this is supportive of our argument that CLTS is more effective at increasing toilet ownership and at decreasing open defecation in relatively poorer areas. 

\begin{figure}
    \centering
    \caption{CLTS treatment effects by night light intensity over study area}\label{fig:scatter}
    \begin{subfigure}[b]{0.45\textwidth}
        \includegraphics[width=\textwidth]{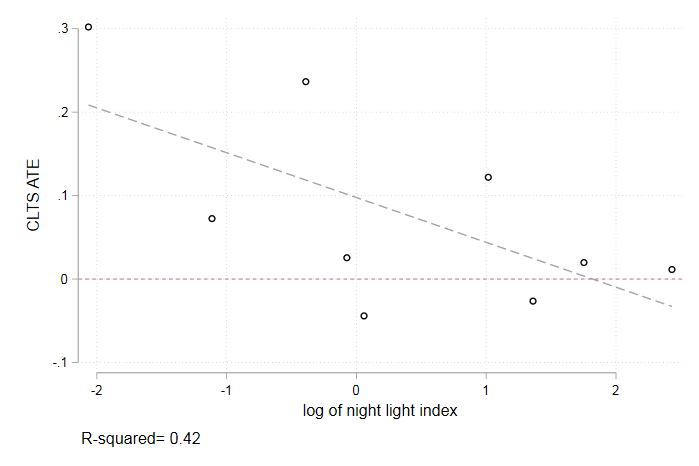}
        \caption{Toilet ownership}
        \label{fig:scatfunclat}
    \end{subfigure}
    ~ 
    \begin{subfigure}[b]{0.45\textwidth}
        \includegraphics[width=\textwidth]{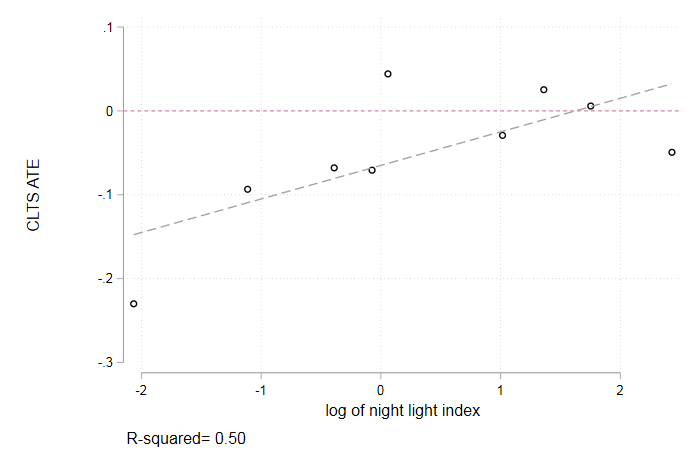}
        \caption{Open defecation}
        \label{fig:scatod}
    \end{subfigure}
    \caption*{\footnotesize \emph{Note:} Figures plot CLTS average treatment effects on toilet ownership (Panel a) and OD rates (Panel b), from each (sub)sample in countries in Figure \ref{f:compare}, except Mali for which we do not have within country variation. The horizontal axis is the logarithm of night light intensity over each study area during baseline.}
\end{figure}

%BRITTA: I WOULD START THIS PARAGRAPH DIFFERENTLY. MY SUGGESTION: We conduct additional analysis to substantiate the inverse relationship between SES and CLTS impacts, shown in Figure 5. For one...
As Figure \ref{f:compare} is suggestive of an inverse relationship between community SES and CLTS impacts, we explore this further by producing a linear fit between the point estimates of CLTS on toilet ownership and open defecation by (the log of) area-specific night light index (see Figure \ref{fig:scatter}). The R-squared of these linear fits suggests that variation in log night light can rationalise 42\% (50\%) of the variation in CLTS effectiveness in increasing toilet ownership (and reducing OD). 

% old text: In a separate exercise, we pooled the five studies for which data was available (i.e. we excluded Mali) and extracted average night light intensity at the second administrative level (districts). We included an interaction term between $T$ and the measure of night light at the district level in equation \ref{eq:compare}, and found that CLTS increases toilet ownership (coefficient 0.092, p-value 0.001) and reduces open defecation (coefficient -0.080, p-value 0.007). As suggested in Figure \ref{f:compare}, the treatment effect is slightly lower in areas with higher night light intensity, as observed in the coefficient on the interaction term, for both toilet ownership (coefficient -0.015, p-value 0.072) and open defecation (coefficient 0.010, p-value 0.220). These results are only suggestive, but are consistent with our findings. A more detailed analysis would require access to the exact cluster locations in each study or more detailed data regarding household ownership of consumer durables (both unavailable), that could then be combined to separately identify cross-study variation from genuine observational differences across households as, for example, proposed by \citet{Meager2018UnderstandingExperiments}. 

In a separate exercise, we pool the three studies for which GPS data was available and we could thus measure nightlight intensity at the cluster level (i.e. Indonesia, Nigeria and Tanzania) and estimate overall and heterogeneous CLTS impacts on toilet ownership, where reporting should be straightforward.\footnote{Please note that when using pooled estimates, we cannot rely on strict exogeneity by randomisation for identification. Yet, we include country fixed effects which will pick up sampling variation across RCT sites, but do not find that they change the impact estimate (see columns 1 and 2 in Table \ref{t:pooled}).} We find qualitatively similar average CLTS impacts to the Nigerian RCT, in the magnitude of a five percentage point increase in toilet ownership (see columns 1 and 2 in Table \ref{t:pooled}). We then estimate heterogeneous impacts for the pooled sample using three alternative functional forms to capture nightlight variation. First, we split geographic units according to whether they display zero or positive nightlight intensity, as its distribution is strongly skewed to the right. Second, we use the results from our Nigerian RCT as reference and define the Nigerian median as a split point as we found no CLTS impact estimates beyond this level in our RCT (see Table \ref{t:tuhte}). Third, we estimate heterogeneous impacts by nightlight intensity using a more flexible split into tertiles. The results, shown in columns 3 to 5 of Table \ref{t:pooled}, support our hypothesis that CLTS impacts vary by communities' SES, here measured through nightlight intensity. We find substantially larger impact estimates in lower SES areas, in the magnitude of 9 percentage points in areas with zero nightlight, i.e. the lowest tertile\footnote{31\% of the observations in the pooled sample are located in areas with zero night light intensity.}. These results get even stronger if we split areas along the Nigerian nightlight median: in poorer areas CLTS increases toilet ownership by 12 percentage points (significant at the 1\% confidence level). Similarly, we never find statistically significant CLTS impact estimates in high SES areas in any specification (be they defined via positive nightlight intensity, intensity above the Nigerian median or in the upper tertile). We further test whether the difference between the estimated coefficients in ``poor'' and ``rich'' areas differ from zero and reject the hypothesis in all specifications but one.\footnote{ The exception is the specification in column 3 where we split the sample into zero and positive nightlight areas. It is likely that this split is too coarse and puts low SES areas with very low but positive nightlights into the high SES category. } Furthermore, impacts are declining across tertiles of increasing nightlight (column 5), similar to the results in our RCT (see the quartile split displayed in Figure \ref{f:tuwealthq}).

\begin{table}[ht]
	\centering
	\begin{threeparttable}
		\caption{Pooled CLTS impacts by nightlight} \label{t:pooled}
		\footnotesize
		\input{content/tableNL_lock_v2.tex}
		\begin{tablenotes}[flushleft]
			\item \emph{Notes:} Pooled regression results using the Indonesian, Nigerian and Tanzanian samples. All specifications control for gender, age and age squared of the household head, as well as a dummy variable equal to one if farming is the main economic activity of the household. District fixed effects are also included and errors are clustered at the level of the randomisation unit.
		\end{tablenotes}
	\end{threeparttable}
\end{table}

These results are only suggestive, but are consistent with our findings. A more detailed analysis would require access to the exact cluster locations in each study or more detailed data regarding household ownership of consumer durables (both unavailable), that could then be combined to separately identify cross-study variation from genuine observational differences across households as, for example, proposed by \citet{Meager2018UnderstandingExperiments}. 

We thus argue that the large range of CLTS impacts across studies can be rationalized by differences in the average wealth (measured through night light) of the area in which they were conducted. Our results - in Nigeria and beyond - suggest that CLTS has been more effective in poorer communities. 

\subsection{Policy Implications}\label{sec:recommendation}

The implications of our findings are self-evident: Governments with restricted funds may achieve larger improvements in sanitation if they use an effective targeting strategy and such a strategy can be based on a measure of community wealth.

We argue that available surveys such as the Demographic and Health Surveys (DHS), or satellite nightlight intensity, both of which are available for all of the 60 countries across continents where CLTS is widely implemented, can form the basis of such a targeting strategy.

\begin{figure}[htbp]
	\centering
	\caption{CLTS targeting in Nigeria}\label{f:mapCLTSNigeria}
		\includegraphics[width=0.5\textwidth]{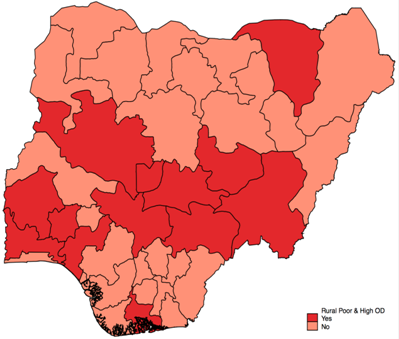}
		\caption*{\emph{Source}: Own calculations based on DHS Nigeria 2013.}
		\label{f:targetting}
\end{figure}

In Appendix \ref{Apsec:dhscomp}, we demonstrate in detail how such data can be used for CLTS targeting in the case of Nigeria. We develop a targeting strategy based on our impact estimates from the Nigerian RCT and the 2013 Nigerian Demographic and Health Survey. Even though the DHS contains a less detailed list of assets than our data, the simpler DHS index of community wealth strongly predicts the more sophisticated measure of community wealth used in our study, supporting the notion that readily available surveys collecting asset wealth information, such as the Demographic and Health Survey, are well-suited for CLTS targeting. The resulting targeting map in Figure \ref{f:mapCLTSNigeria} highlights priority areas for targeting, i.e. poor areas where toilet ownership rates are low, in darker shaded areas.

%% file: content/tableNL_lock_v2.tex
\begin{tabular}{l*{5}{c}}
\hline\hline
Dep. variable: toilet ownership& \multicolumn{2}{c}{All} & \multicolumn{3}{c}{By area nightlight intensity (NL)}\\
[1em]
&&& \multicolumn{1}{c}{Zero vs. pos.} & \multicolumn{1}{c}{Below vs. above} & \multicolumn{1}{c}{NL tertiles}\\
&&& \multicolumn{1}{c}{nightlight} & \multicolumn{1}{c}{Nigerian median} & \\
\cmidrule(lr){2-3}\cmidrule(lr){4-4}\cmidrule(lr){5-5} \cmidrule(lr){6-6}
                    &\multicolumn{1}{c}{(1)}&\multicolumn{1}{c}{(2)}&\multicolumn{1}{c}{(3)}&\multicolumn{1}{c}{(4)}&\multicolumn{1}{c}{(5)}\\

\hline
\multicolumn{5}{l}{\textit{Pooled average impact}}\\
[1em]
CLTS                &        0.05&        0.05&            &             &       \\
                    &      (0.01)&      (0.01)&            &             &      \\
[1em]
\hline
\multicolumn{5}{l}{\textit{Heterogeneity by communities' nightlight intensity}}  \\
[1em]
CLTS in zero/below median/1st tertile NL          &            &           &        0.09&          0.12&     0.09\\
					&            &           &      (0.01)&        (0.00)&     (0.01)\\
[1em]
CLTS in 2nd tertile NL      &            &           &            &            &        0.07\\
                    &            &           &            &            &    (0.04)\\
[1em]
CLTS in positive/above median/3rd tertile NL          &            &           &        0.03&         -0.01&      -0.01\\
					&            &           &      (0.25)&        (0.62)&      (0.69)\\
[1em]
Difference          &            &           &       -0.06&         -0.13&            \\
					&            &           &      (0.14)&        (0.00)&            \\
[1em]
Difference (Middle-Low)&         &           &            &            &       -0.02\\
                    &            &           &            &            &       (0.71)\\
[1em]
Difference (High-Low)&           &           &            &            &        -0.10\\
                    &            &           &            &            &       (0.03)\\
\hline
Country FE          &          No&         Yes&        Yes&         Yes&         Yes\\
BL ownership        &          No&          No&         No&          No&          No\\
No. of clusters     &         580&         580&        580&         580&         580\\
No. of obs.         &       7,843&       7,843&      7,843&       7,843&       7,843\\
\hline\hline
\end{tabular}

%% file: 8.conclusion.tex
\section{Discussion and Conclusion}\label{sec:conclusion}

The design of effective policies to address the urgent sanitation concerns in the developing world requires a nuanced understanding of households' investment choices and drivers of behavioral change. In this paper we provide evidence on the effectiveness of Community-Led Total Sanitation (CLTS), a participatory information intervention widely implemented around the world.

Our study uses a large cluster randomized experiment in Nigeria for which we collected data up to three years after treatment. Implementation of CLTS was conducted at-scale, i.e. by WASH civil servants trained by local NGOs. We show that CLTS, a participatory community intervention without financial components, had positive but moderate effects on open defecation and toilet construction overall. However, average impacts hide important heterogeneity by communities' socio-economic status, as the intervention has strong and lasting effects on open defecation habits in poorer communities, and increases sanitation investments. In poor communities, OD rates decreased by 9pp from a baseline level of 75\%, while we find no effect in richer communities. The reduction in OD is achieved mainly through increased toilet ownership (+8pp from a baseline level of 24\%). While this result is robust across several measures of community socio-economic status, and is not driven by baseline differences in toilet coverage, our data does not allow us to pin down why households in poorer communities are more susceptible to the programme. However, in addition to the more effective targeting strategies, highlighted in the previous section, that governments can adopt, our results have three further important implications.

First, our results provide an additional reason why scale-up of interventions is not trivial \citep{Ravallion2012,BoldEtAl2013,BanerjeeEtAl2017,DeatonCartwright2018}. Discussions on why interventions may not scale-up successfully in a national roll-out have focused on general equilibrium and spillover effects, and recently on aspects of implementation and delivery. The literature has suggested that spillovers and moderating general equilibrium effects may lead to lower returns to interventions, when interventions conducted in areas with specific characteristics are being rolled out universally, e.g. in richer areas. We show that community-specific, heterogeneous treatment impacts are an additional impediment to successful scale-up in terms of effectiveness of interventions.

Second, community SES also provides plausible external validity beyond our Nigerian-based RCT. Using data from our study and five other RCTs of similar interventions, we find an inverse relationship between area-level wealth, measured by night light intensity, and program effectiveness across these studies. Thus, we have identified a characteristic that rationalizes the wide range of impact estimates in the literature. 

Last but not least, we show that interventions relying on information and collective action mechanisms can have substantial impacts on households' health investments and behaviour, specifically relating to sanitation. Yet , there is an important caveat for policy-makers working towards meeting the sanitation-related sustainable development goals. CLTS achieves convergence between poor and rich communities in terms of OD and toilet coverage in our study - and thus levels the playing field. However, it is not a silver bullet to achieve open defecation \emph{free} status in poor communities. Hence, more research on alternative or supplementary interventions to close the sanitation gap in low income countries is needed. These may either seek to magnify CLTS impacts (e.g. through complementary financial incentives, loans or subsidies or more intensive followup), or improve sanitation in rich communities where CLTS is ineffective, e.g. via infrastructure investment and supply side interventions.

%\author{Laura Abramovsky\thanks{Centre for the Evaluation of Social Policies (EDePo), Institute for Fiscal Studies} \and Britta Augsburg\footnotemark[2] \and Melanie L\"uhrmann\thanks{Department of Economics, Royal Holloway, and IFS} \and Francisco Oteiza\thanks{Institute of Education, University College London} \and Juan Pablo Rud\footnotemark[3]}

\vspace{10mm}
\begin{singlespace}
\noindent THE INSTITUTE FOR FISCAL STUDIES 

\noindent THE INSTITUTE FOR FISCAL STUDIES 

\noindent ROYAL HOLLOWAY AND THE INSTITUTE FOR FISCAL STUDIES

\noindent INSTITUTE OF EDUCATION, UNIVERSITY COLLEGE LONDON 

\noindent ROYAL HOLLOWAY AND THE INSTITUTE FOR FISCAL STUDIES
\end{singlespace}

%% file: 9a.Appendix.tex
\newpage 
\begin{appendix}
\section*{Appendix}

%%%%%%%%%%%%%%%%%%%%%%%%%%%%%%%%%%%%%%%%%%%%%%%%%%%%%%%%%%%%%%%%%%%%%%%%%%%%%
%%% VARIABLE DEFINITIONS
%%%%%%%%%%%%%%%%%%%%%%%%%%%%%%%%%%%%%%%%%%%%%%%%%%%%%%%%%%%%%%%%%%%%%%%%%%%%%

\section{Variable definitions}\label{Apsec:vardesc}

In this section, we provide details on a series of measurements used to construct household and community-level characteristics. These are based on our household surveys and other auxiliary datasets. 

%%%%%%%%%%%%%%%%%%%%%%%%%%%%%%%%%%%%%%%%%%%%%%%%%%%%%%%%%%%%%%%%%%%%%%%%%%%%%%%%%%%%%%%%%%%%%%%%%%%%%%%%%%
%%%%%%%% HOUSEHOLD CHARACTERISTICS
%%%%%%%%%%%%%%%%%%%%%%%%%%%%%%%%%%%%%%%%%%%%%%%%%%%%%%%%%%%%%%%%%%%%%%%%%%%%%%%%%%%%%%%%%%%%%%%%%%%%%%%%%%

\subsection{Household characteristics}\label{Apsubsec:hhchars}

\noindent\emph{Asset wealth}\\
Household survey measures of annual household income had relatively low response rates: 27\% of the households interviewed reported no income at all or refused to answer. A higher response rate was achieved in a list of questions regarding the ownership of consumer durables. We applied a principal component analysis to this list, and constructed an index of asset wealth based on the first principal component, following \cite{Filmer2001EstimatingIndia}. Such weatlh indices are widely used as a proxy for household long-term wealth, for example, in the USAID Demographic and Health Surveys (DHS) run over 90 countries, or as a targeting tool for the PROGRESA conditional cash-transfer programme \citep{mckenzie2005measuring}. Table \ref{t:rwi} lists the asset items elicited in our household survey, and shows their factor loadings. 

% Variables included in the RWI index
\begin{table}[ht] 
    \centering
    \begin{threeparttable}
    \caption{Asset items used in the asset wealth index} \label{t:rwi}
    \footnotesize
    \begin{tabular}{ l  c  }
    \hline
    \textbf{Survey question}  & \textbf{Factor loading}  \\ 
    \hline 
    \textit{Ownership of the following durable assets: (1=Yes, 0=No)}& \\
    Motorcycle/scooter/tricycle & 0.1302 \\
    Furniture: chairs & 0.1561  \\
    Furniture: tables & 0.1823  \\
    Furniture: beds & 0.1075    \\
    Refrigerator & 0.2998       \\
    Washing machine & 0.1826    \\
    Microwave oven & 0.1914     \\
    Gas cooker & 0.2507         \\
    Plasma/flat screen TV & 0.2173 \\
    Other TV & 0.2867           \\
    Satellite dish (monthly subscription) & 0.2272 \\
    Other satellite dish (DSTV, etc) & 0.2391 \\
    Radio/CD/DVD Player & 0.2241 \\
    Smart phones  & 0.1265       \\
    Other Telephone / phones & 0.0886 \\
    Computer & 0.2195\\
    Air conditioner & 0.1061\\
    Power generator & 0.2777\\
    Sewing machine & 0.1323\\
    Electric iron & 0.3172\\
    Pressure cooker & 0.1557\\
    Electric fans & 0.3162\\
    \hline
    Number of households included (N=4,722) & 4,622 \\
    \hline
    \end{tabular}
    \begin{tablenotes}[flushleft]
			\item \emph{Notes:} Questions were coded equal to one if the household reported to own at least one of each of the items listed in each category. The wealth index was then constructed using the first component of the principal component analysis. Households with missing data for at least one of the categories were excluded.
		\end{tablenotes}
    \end{threeparttable}
\end{table}

%%%%%%%%%%%%%%%%%%%%%%%%%%%%%%%%%%%%%%%%%%%%%%%%%%%%%%%%%%%%%%%%%%%%%%%%%%%%%%%%%%%%%%%%%%%%%%%%%%%%%
%%%%%%%% COMMUNITY CHARACTERISTICS
%%%%%%%%%%%%%%%%%%%%%%%%%%%%%%%%%%%%%%%%%%%%%%%%%%%%%%%%%%%%%%%%%%%%%%%%%%%%%%%%%%%%%%%%%%%%%%%%%%%%%

\subsection{Community characteristics}\label{Apsubsec:commchars}

A community is on average composed of one to two villages or neighborhoods, consisting of 220 households (see details in Section \ref{sec:design}).\bigskip

%\subsubsection{Socio-economic conditions}
\noindent\emph{Community wealth}\\
Community asset wealth is estimated as the median household's asset wealth score. Our household survey randomly interviewed 20 households per community, so we chose the median to limit possible distortions due to outliers (i.e. households with extremely high or low asset wealth). Our main results discretise community wealth along the median. Poor (rich) communities were those with asset wealth below (above or equal) the median community. 

In addition to community wealth, we propose three alternative measures of communities' socio-economic status: a night light index, population density, and isolation.\bigskip

\noindent \emph{Night light index}\\
The first alternative measure is the average night light index recorded in 2013, before the intervention began. Using household GPS coordinates, we calculate the geographical centroid of each community, and define a 5km radius around the centoid. We used nighttime lights data made available by the U.S. National Oceanographic and Atmospheric Administration (NOAA). The observations on which the data is assembled are made by the Operational Linescan System (OLS) flown on the Defense Meteorological Satellite Program (DMSP) satellites.

The intensity of nighttime lighting has been shown to be correlated with economic growth, and proposed as a tool for inferring growth rates at the sub-national level by \cite{Henderson2012MeasuringSpace}. Moreover, \cite{Michalopoulos2013} show that nighttime light intensity is highly correlated with GDP per capita and urbanization across African countries. \bigskip

\noindent \emph{Isolation}\\
Communities' geographical isolation is measured using GIS software. We calculate the linear distance, measured in kilometers, between each community centroid and the nearest administrative capital. Nigerian states are divided into local government areas (LGAs), and each LGA has its own administrative headquarters. Note that we measure the distance to the nearest LGA capital, which may or may not be the capital of the LGA in which a community is located. Results are unchanged when we use the distance to the nearest \textit{state} capital as the relevant administrative capital instead. \bigskip

\noindent\emph{Population density}\\
We compute the number of households living within the 5km radius radius around each community centroid. We rely on a census of households living in the area during mid 2014, conducted in preparation for the evaluation. This acts as a measure of population density, as the radius area is equal for all communities ($\pi \times 5^2 =  79 km^2$). 

As for asset wealth, we split the communities in our sample into two groups along each of these measures, creating two (approximately) identically sized sub-samples along each dimension.  \bigskip

\noindent \emph{Toilet coverage}\\
Toilet coverage is the share of households in the community that owned toilets at baseline. \bigskip

% \subsubsection{Additional community characteristics}

\noindent \emph{Religious fragmentation}\\
Religious fragmentation denotes the probability that two randomly selected households in our sample are of different religions. We focus on religious rather than ethnic fragmentation, as our study sample is extremely homogeneous along ethnic lines but exhibits considerable religious diversity.\footnote{ Similarly defined indices, first used in \cite{Mauro1995CorruptionGrowth}, are frequently used to study ethnic diversity and its impact on economic growth \citep{Easterly1997AfricasDivisions} or public goods provision \citep{Alesina1999PublicDivisions}.} \bigskip

\noindent \emph{Trust}\\
In a seminal paper, \cite{Alesina2002WhoOthers} define trust based on a question from the US General Social Survey, asking respondents if they thought that `most people can be trusted'. Since we want to uncover trust \emph{within} the community rather than a concept of social trust, we instead rely on the following question: `Generally speaking, would you say that you trust the people in your neighborhood a lot, only a little, or not at all?'. Responses were coded with 2 (a lot), 1 (a little) or 0 (not at all). We use mean values of household responses within a community as community-level aggregate.\bigskip

\noindent \emph{Social capital}\\
An index of household level social capital was constructed by principal component analysis using questions related to participation in community life, social contact and provision of services to the community (see Table \ref{t:social}). It is similar to the one used in \cite{Cameron2019} who study CLTS impacts in the province of East Java, Indonesia. Table \ref{t:social} shows each item's factor loadings. We aggregate household responses to a community-level measure by taking the community-specific mean.\bigskip

% Variables included in the SC index
\begin{table}[ht] 
    \centering
    \begin{threeparttable}
    \caption{Questions used to construct the household-level social capital index} \label{t:social}
    \footnotesize
    \begin{tabular}{ l  c }
    \hline
    \textbf{Survey question}  & \textbf{Factor loading} \\ 
    \hline 
    \textit{How many times in the past 12 months have you ...} & \\
    ... donated blood? & 0.0573\\
    ...worked on a community project?  & 0.3129\\
    ...attended any public meeting in which there was discussion of town or school affairs?  & 0.2939\\
    ...attended a political meeting or rally? & 0.2387 \\
    ...attended any club or organizational meeting (not including meetings for work)? & 0.3204 \\
    ...had friends over to your home?  & 0.3662 \\
    ...been in the home of/invited a friend of a different race?  & 0.3166 \\
    ...been in the home of/invited someone of a different neighbourhood?  & 0.3573 \\
    ...been in the home of/invited someone you consider to be a community leader?  & 0.3699\\
    ...volunteered?  & 0.3185\\
    ...served as an official or served on a committee of a club or community association?  & 0.2259 \\
    Not including weddings and funerals, how often do you attend religious services?  & 0.0382 \\
    \hline
    Number of households included (N=4,722) & 4,227 \\
    \hline
    \end{tabular}
    \begin{tablenotes}[flushleft]
			\item \emph{Notes:} These were multiple choice questions in which the (pre-specified) answers ranged from `Never did this' to `More than once a week'. For the purposes of constructing the social capital index, these responses were standardized, before conducting the principal component analysis. Households with missing data for at least one of the categories were excluded.
		\end{tablenotes}
    \end{threeparttable}
\end{table}

\noindent \emph{Wealth inequality}\\
Wealth inequality is constructed by dividing the \textit{within community} standard deviation of household asset wealth by the standard deviation of household wealth over the whole sample. \cite{mckenzie2005measuring} shows that, in the absence of reliable information on household income or consumption, this measure of asset wealth inequality is an informative proxy for inequality in living standards.  \bigskip

\begin{figure}[ht]
	\centering
	\caption{Community characteristics at baseline} \label{f:histograms}
	\includegraphics[width=\textwidth]{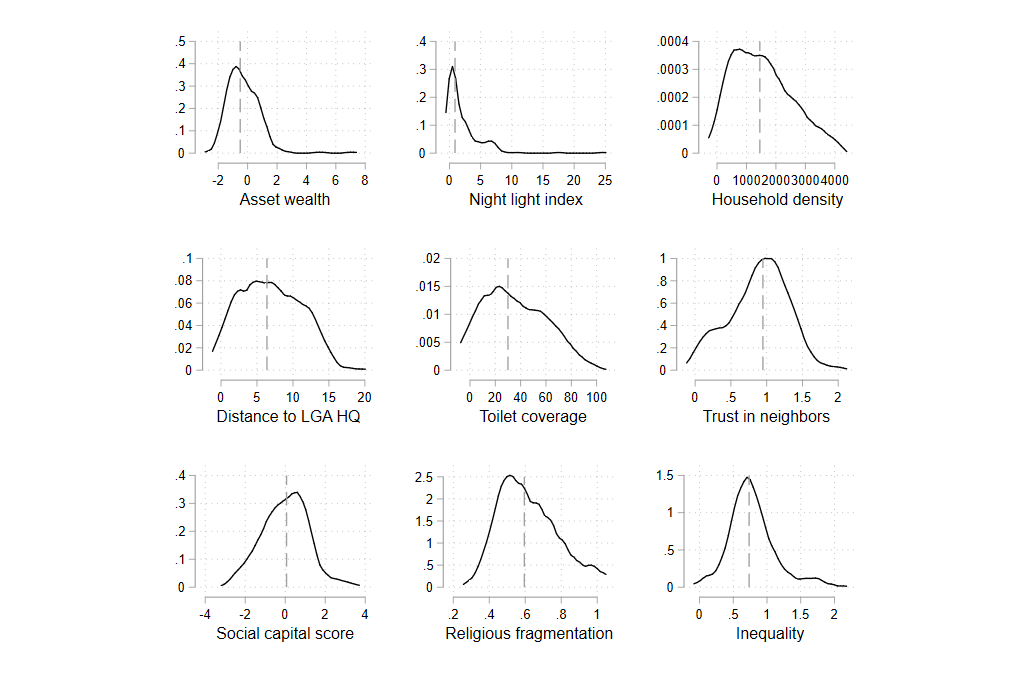}
	\caption*{\emph{Note:} Kernel density approximations of the distribution of communities in our sample, according to a series of community level characteristics. Dashed lines indicate median values, along which the sample was split to arrive at groups with high and low scores for each trait.}
\end{figure}

\noindent \emph{Distribution of key community characteristics}\\
Figure \ref{f:histograms} shows the distribution of each of the community characteristics described above. The strong skewness in in night light intensity across communities is one reason for the discrete measures we use in our preferred specification (and for the sensitivity analysis using the continuous measures). %The Figure shows clearly that a specification that interacts treatment status with a continuous measure of night light intensity does not represent the underlying distribution of night light intensity values appropriately. 

% Variation of TU characteristics along which we estimate CLTS HTEs
\bigskip

\noindent \emph{Communities' public infrastructure}\\
Local school or hospital infrastructure as well as an indicator whether the community has paved internal roads serve as further measures of local development status. These measures come from a community level questionnaire conducted at baseline, posed to community leaders.\bigskip 

\noindent \emph{Characteristics of village leaders}\\
CLTS is triggered after an initial meeting with the village leader who helps organize the community meeting, and aims to spurt collective action towards an open defecation free status of the community. Hence, village leaders may influence their communities' characteristics. As a community may comprise several villages, these characteristics reflect their mean age, experience (years in leadership), and education attainment, all measured at baseline.\bigskip

%% file: 9b.Appendix.tex
\newpage
%%%%%%%%%%%%%%%%%%%%%%%%%%%%%%%%%%%%%%%%%%%%%%%%%%%%%%%%%%%%%%%%%%%%%%%%%%%%%
%%% Attrition
%%%%%%%%%%%%%%%%%%%%%%%%%%%%%%%%%%%%%%%%%%%%%%%%%%%%%%%%%%%%%%%%%%%%%%%%%%%%%
%\section{Household survey attrition}\label{Apsec:attrition}
\section{Attrition, compliance and implementation}\label{Apsubsec:attrition}

We provide additional results on survey attrition, on non-compliance of communities with their treatment assignment, and consider implementation differences across the two study states. 

\bigskip \noindent \emph{Attrition}\\
First, we test for non-random attrition from household surveys across the three followup studies in treatment and control communities. Our study showed remarkably low attrition rates: 2.53\% in FU1, 8.81\% in FU2 and 11.58\% in FU3 (Table \ref{t:att1}). Secondly, attrition rates were similar across treatment arms. Third, we find slightly lower attrition rates in poor communities than in rich ones.

\begin{table}[ht]
	\centering
	\begin{threeparttable}
		\caption{Attrition rates by survey wave and treatment status} \label{t:att1}
		\footnotesize
		\input{content/att1.tex}
		\begin{tablenotes}[flushleft]
			\item \emph{Notes:} Unconditional attrition rates, as a share of the total attempted interviews in each survey wave.
		\end{tablenotes}
	\end{threeparttable}
\end{table}

In Table \ref{t:att2} we formally test whether treatment status can predict attrition conditional on baseline characteristics. Columns 1, 4 and 7 show the results of a regression of attrition on the treatment indicator and district fixed effects, for FU1, FU2, and FU3 respectively. We see that assignment to CLTS does not predict attrition, and this results is robust to the inclusion of household-level controls (columns 2, 5 and 8). Hence, we find no evidence of selective attrition that would challenge the successful randomization demonstrated in Table \ref{t:balanceall}. 
In FU 1 and FU2, attrition is balanced not only across treatment groups, but also by community wealth. Only in FU3, we find slightly higher attrition rates (3pp) among rich communities than poor ones (p-value=0.02). Yet, recall that even three years after the baseline survey, mean attrition is with around 10\% very low. 

\begin{table}[ht]
	\centering
	\begin{threeparttable}
		\caption{Tests for non-random attrition} \label{t:att2}
		\footnotesize
		\input{content/att2.tex}
		\begin{tablenotes}[flushleft]
			\item \emph{Notes:} Estimation results for regressions of attrition, by wave, on treatment status, household and community level characteristics. Errors are clustered at the community level and \textit{p}-values shown in parenthesis.
		\end{tablenotes}
	\end{threeparttable}
\end{table}

\bigskip \noindent \emph{Implementation and Triggering}\\
We test if differences in the quality of CLTS delivery may explain heterogeneous impacts between poor an rich communities. As described in Section 2.2, WaterAid hired two NGOs with CLTS experience, one from each state, to train local government officials in the facilitation of CLTS meetings. We thus test for state-level differences in CLTS effectiveness. The interaction term is small and not significantly different from zero (p-value=0.326), so we do not find evidence for this hypothesis.

\begin{table}[ht]
	\centering
	\begin{threeparttable}
		\caption{CLTS impacts by state and triggering status} \label{t:implementation}
		\footnotesize
		\input{content/table_implementation_lock.tex}
		\begin{tablenotes}[flushleft]
			\item \emph{Notes:} All specifications control for the household and household head characteristics listed in Table \ref{t:baseline}. Errors are clustered at the community level. Naive (unadjusted) p-values shown in parenthesis.  
		\end{tablenotes}
	\end{threeparttable}
\end{table}

Finally, not all clusters assigned to CLTS were triggered, leading to a fraction of non-complying clusters. Our main results are thus based on unbiased ITT estimates. In column 2 of Table \ref{t:implementation}, we follow \cite{Imbens1994IdentificationEffects} and \cite{Angrist1995Two-stageIntensity}, redefine the treatment variable to reflect actual triggering status, and instrument triggered treatment with treatment assignment. The results on average impact are very similar to the ITT estimates. In columns 3 and 4, we additionally test if triggered clusters in poor communities may be more reactive to CLTS than in rich communities. The results are very similar to the ITT estimates. We find CLTS to be only effective in poor communities, where they reduce OD by 10pp.

%%%%%%%%%%%%%%%%%%%%%%%%%%%%%%%%%%%%%%%%%%%%%%%%%%%%%%%%%%%%%%%%%%%%%%%%%%%%%%%%%%%%%%%%%%%%%
%%%%%%%% SENSITIVITY CHECKS: WHY COMMUNITY WEALTH
%%%%%%%%%%%%%%%%%%%%%%%%%%%%%%%%%%%%%%%%%%%%%%%%%%%%%%%%%%%%%%%%%%%%%%%%%%%%%%%%%%%%%%%%%%%%%

\section{Additional results: heterogeneous impacts}\label{Apsec:het}

\subsection{Alternative specifications of community characteristics}\label{Apsubsec:hetcont}

First, we present sensitivity analysis based on alternative functional forms of the four community characteristics. Instead of discretizing measures along their median cutoffs, as presented in the main text in Table \ref{t:tuhte}, we use continuous measures - with very similar results as Table \ref{t:tuhtecont} shows. Poor treated communities whose wealth is one standard deviation below the mean of zero display a reduction in OD by 10\%, while the two coefficients level out to a close to zero impact for rich CLTS communities. Similar results are found for density and isolation. Similar to the discrete measures, once we correct for family wise errors, we find statistically significant heterogeneous impacts only for community wealth and household density. 

In contrast to the results in Table \ref{t:tuhte}, we find no heterogeneous impact in the continuous nightlight measure. Figure \ref{f:histograms} shows why. In contrast to the other three characteristics, its distribution is highly skewed and displays strong bunching around zero for about 90 out of 247 communities. Hence, a discrete specification seems more suited to capture heterogeneity in nightlight intensity. 

%CLTS HTE by COMMUNITY SES PROXIES - CONTINUOUS
\begin{table}[ht]
	\centering
	\begin{threeparttable}
		\caption{CLTS impacts on OD using four measures of community SES} \label{t:tuhtecont}
		\footnotesize
		\input{content/tableTUHTE_OD_2_CONT.tex}
		\begin{tablenotes}[flushleft]
			\item \emph{Notes:} All specifications control for the household and household head characteristics listed in Table \ref{t:baseline}. Errors are clustered at the community level. Naive (unadjusted) p-values shown in parenthesis. In brackets we present p-values adjusted by family wise error rate following \cite{romano2005stepwise}, using 1,000 cluster bootstrap samples.
		\end{tablenotes}
	\end{threeparttable}
\end{table}

Secondly, we run a horserace between our four (imperfectly) correlated measures of communities' socio-economic conditions to determine which one best captures heterogeneous impacts across communities.
Columns 1 to 3 in Table \ref{t:SESrobust} show estimates allowing for pairwise heterogeneity (i.e. by community wealth and one of the other three variables). We find heterogeneous impacts with respect to all three alternative characteristics. This suggests that each SES indicator captures a community aspect which matters for the successful implementation of CLTS. Yet, the most striking result is this: We detect strongly statistically significant and economically identical CLTS impacts in poor communities (that are also of equal magnitude than those estimated in Table \ref{t:tuhte}) in all specifications. In fact, when we include all interaction terms simultaneously (column 4), only the community wealth-specific impacts are statistically significant at the 5\% level.

% Horserace between the different SES measures
\begin{table}[ht]
	\centering
	\begin{threeparttable}
		\caption{Performance of different proxies for community-level socio-economic status (SES)} \label{t:SESrobust}
		\footnotesize
		\input{content/tableSESrobust.tex}
		\begin{tablenotes}[flushleft]
			\item \emph{Notes:} All specifications control for the household and household head characteristics listed in Table \ref{t:baseline}. \emph{p}-values are shown in parenthesis. Standard errors are clustered at the community level. 
		\end{tablenotes}
	\end{threeparttable}
\end{table}

These results underline the robustness of our finding that CLTS is effective only in poor communities, and its effectiveness is decreasing in the wealth of the targeted community. The horserace exercise suggests that community wealth is the best proxy for local socio-economic conditions in a community, and plays an important role in mediating CLTS impacts.\\

%Table \ref{t:tuhte2cont} analogously shows estimates based on the continuous measures of fragmentation, social capital and trust. Again, the results are robust to both functional form choices. 

%CLTS HTE by OTHER COMMUNITY CHARACTERISTICS - CONTINUOUS
%\begin{table}[ht]
%	\centering
%	\begin{threeparttable}
%		\caption{CLTS impacts on OD by community characteristics} \label{t:tuhte2cont}
%		\footnotesize
%		\input{content/tableTUHTE_OD_CONT.tex}
%		\begin{tablenotes}[flushleft]
%			\item \emph{Notes:} All specifications control for the household and household head characteristics listed in Table \ref{t:baseline}. Errors are clustered at the community level. Naive (unadjusted) p-values shown in parenthesis. In brackets we present p-values adjusted by family wise error rate following \cite{romano2005stepwise}, using 1,000 cluster bootstrap samples and estimated jointly for the regressions presented in Tables \ref{t:tuhtecont} and \ref{t:tuhte2cont}. 
%		\end{tablenotes}
%	\end{threeparttable}
%\end{table}
%}

\subsection{Are heterogeneous CLTS driven by community or household wealth?}\label{Apsubsec:commvshh}

Community wealth is an aggregated measure of household wealth. Richer (poorer) communities are on average composed of richer (poorer) households.\footnote{ 69\% of the households living in poor clusters have below-median asset wealth, while 66\% of the households living in rich clusters have above-median wealth.} So is CLTS is simply (more) effective among poorer households? We test this hypothesis in two ways. First, we show that CLTS is more effective among poorer households (see column 1 of Table \ref{t:wealthhhvscomm}). We find no statistically significant CLTS impact among rich households, while poor households exhibit a 5pp decline in OD induced by CLTS. Yet, the impact of CLTS in poor households is much smaller than the 10pp decline in OD detected in poor communities in column 1 in Table \ref{t:tuhte}. 

% Alternative specification for HH wealth variable
\begin{table}[ht]
	\centering
	\begin{threeparttable}
		\caption{CLTS impacts by household level wealth} \label{t:wealthhhvscomm}
		\footnotesize
		\input{content/table_wealth_lock.tex}
		\begin{tablenotes}[flushleft]
			\item \emph{Notes:} All specifications control for the household and household head characteristics listed in Table \ref{t:baseline}. Errors are clustered at the community level. Naive (unadjusted) p-values shown in parenthesis. In brackets we present p-values adjusted by family wise error rate following \cite{romano2005stepwise}, using 1,000 cluster bootstrap samples and estimated jointly for all regressions presented in Tables \ref{t:wealthhhvscomm} and \ref{t:hhhte}.
		\end{tablenotes}
	\end{threeparttable}
\end{table}

Secondly, if CLTS impacts are heterogeneous across households rather than communities, we would expect to find stronger CLTS impacts (in effect size and statistical significance) among poor households in either community type. Split households along the sample median into poor and rich, and separately estimating heterogeneous CLTS impacts by household wealth in rich and poor communities, we reject this hypothesis (columns 2 and 3). Instead, we find statistically significant impacts of CLTS for poor and rich households, but only in poor communities. Additionally, there is no evidence of differential CLTS impacts among poor households in either community type.

Aggregating to community level may reduce measurement error in household wealth, and thus provide a more robust measure of households' SES, or long term wealth. To address this concern, we additionally used (completed primary) education of the household head as a proxy for household wealth, and find no evidence of heterogeneous impacts along this dimension (see column 1 in Table \ref{t:hhhte}).\\

\begin{table}[ht]
	\centering
	\begin{threeparttable}
		\caption{CLTS impacts on OD by household characteristics} \label{t:hhhte}
		\footnotesize
		\input{content/tableHHHTE2_OD_lock.tex}
		\begin{tablenotes}[flushleft]
			\item \emph{Notes:} All specifications control for age and employment status of the household head, as well as household size. Errors are clustered at the community level and adjusted by family wise error rate following \cite{romano2005stepwise}.  
		\end{tablenotes}
	\end{threeparttable}
\end{table}

\noindent \emph{Other household characteristics}\\
It is often posited that women and households with children may exhibit a higher willingness to invest into health and sanitation technologies.\footnote{ With respect to CLTS, \cite{kar2003} poses that women are `one of the greatest internal forces for mobilisation and promotional activities in the villages'. Women may enjoy larger returns from private sanitation in terms of personal safety and privacy. Evidence from other health enhancing investments suggests the existence gender-specific preferences in certain domains, such as health and children's welfare \citep{miller2013gender}.} In columns 2 and 3 of Table \ref{t:hhhte}, we test these hypotheses using indicators for female headed households and those with children, and find no evidence of heterogeneous CLTS impacts in either of these dimensions.

\subsection{Sensitivity analysis of CLTS impacts on sanitation investments}\label{Apsubsec:channels}

We re-estimate Table \ref{t:margins}, replacing community wealth by the other three measures of socio-economic conditions. CLTS increases (functioning and general) toilet ownership in communities with low nightlight intensity (see top panel in Table \ref{t:channels_nl}), low density (second panel) and high isolation (third panel). Using these alternative community SES measures, we find no CLTS impact in rich communities in terms of sanitation investments. 

% Channels of OD reduction - by NL
\begin{table}[ht]
	\centering
	\begin{threeparttable}
		\caption{Channels of OD reduction} \label{t:channels_nl}
		\footnotesize
		\input{content/table_channels_TUnlL_lock.tex}
		\begin{tablenotes}[flushleft]
			\item \emph{Notes:} All specifications control for the household and household head characteristics listed in Table \ref{t:baseline}. \emph{p}-values are shown in parenthesis. Standard errors are clustered at the community level and are adjusted for family-wise error rate following \cite{romano2005stepwise}, using 1,000 cluster bootstrap samples. 
		\end{tablenotes}
	\end{threeparttable}
\end{table}

Equally in line with the results presented in Table \ref{t:margins} in the main text, we find little evidence of behavioural change in terms of usage of existing toilets, or shared usage. There is weak indication that shared toilet use increases in areas that are close to the LGA's capital, have high nightlight intensity and are more densely populated. Yet, the differential impact of CLTS across community groups along shared usage is never statistically significant.

%%%%%%%%%%%%%%%%%%%%%%%%%%%%%%%%%%%%%%%%%%%%%%%%%%%%%%%%%%%%%%%%%%%%%%%%%%%%%%%%%%%%%%%%%%%%%%%%%%%%%%%%%%
%%%%%%%% COMPARISON WITH DHS
%%%%%%%%%%%%%%%%%%%%%%%%%%%%%%%%%%%%%%%%%%%%%%%%%%%%%%%%%%%%%%%%%%%%%%%%%%%%%%%%%%%%%%%%%%%%%%%%%%%%%%%%%%

\section{Community wealth measures in household surveys in developing countries}\label{Apsec:dhscomp}

A similar community wealth index to the one in our study sample is widely available in the demographic and household surveys (DHS). These are nationally representative samples conducted in 90 developing countries. Using the 2013 Nigerian Demographic and Health Survey (DHS), we show that a community wealth index can be easily constructed for a national CLTS targeting strategy, and that it precision is similar to the measure in our study data.

%\subsection{Constructing a common wealth index across samples}
We first construct a new, comparable asset wealth index (the `DHS index') in both samples, our study and in the DHS. It is based on the joint subset of durable asset items recorded in both surveys.\footnote{ Asset items elicited in both surveys are: bicycles, motorcycles (scooters, tricycles, etc), cars and trucks, refrigerators, radio, TV, bank account, telephone (mobile or fixed), improved water, improved sanitation and livestock (cattle, goats and sheep, pigs, poultry). The inclusion of livestock is motivated by the purpose of capturing farming households. Indeed, the loading factors for these four elements are negative.} To ensure national representativeness, we construct the index by performing a principal component analysis of the questions on the DHS sample only. The resulting index in our study data is created by applying the estimated factor loadings to households' responses. 

\begin{figure}[ht]
	\centering
	\caption{Household asset wealth indices from our study data are highly consistent} \label{t:hhwealth_newold}	\includegraphics[width=.9\textwidth]{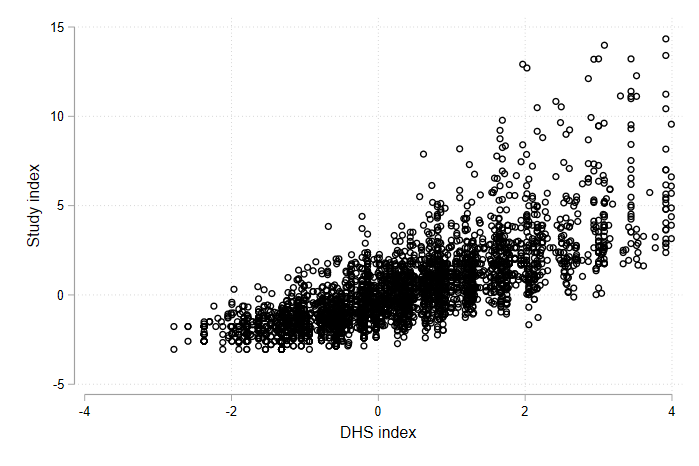}
	\caption*{\emph{Note:} Scatter plot showing the asset wealth scores obtained for each household in our study sample. Each dot represents one household. The y-axis shows the value for the asset wealth index used in our study, built using asset ownership questions from our household survey (see Table \ref{t:rwi}). The x-axis shows the score obtained in the asset wealth index constructed using only those questions that were included in both our household survey and in the DHS, for the same sample of households. See text for more details.}
\end{figure}

In our study sample, the simplified DHS index, based on factor loadings from the DHS sample, is closely matched with the original household wealth index (henceforth: the `Study index'), which was based on a more comprehensive list of asset items (see Table \ref{t:rwi}). Figure \ref{t:hhwealth_newold} shows that the two measures are highly positively correlated ($\rho$=0.77, significant at the 1\% level). The newly created DHS index explains up to 58\% of the variation in the more comprehensive Study index. %In other words, households' wealth is similar, regardless of whether we use a more detailed or a simplified asset wealth index.

%\subsection{Putting our sample into context}

\begin{figure}[ht]
	\centering
	\caption{Distribution of the DHS wealth index by sample} \label{f:hhwealth_dist}	\includegraphics[width=.9\textwidth]{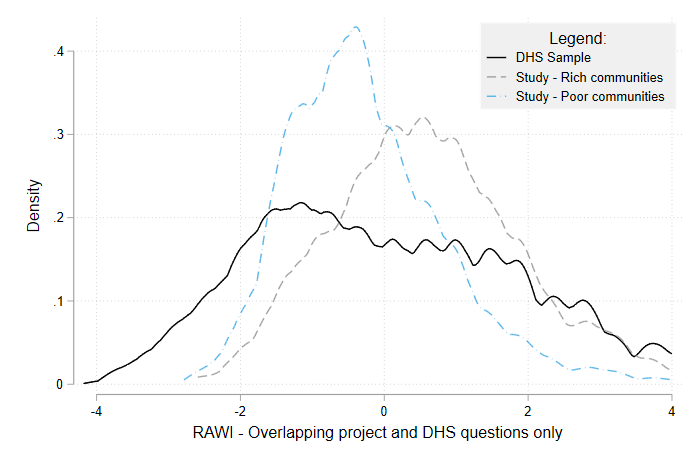}
	\caption*{\emph{Note:} Kernel density plot showing the distribution of DHS index scores among the DHS sample (black), rich (gray) and poor (blue) communities in our study sample. The DHS index is the asset wealth index constructed using only those questions that were included in both our household survey and the DHS.}
\end{figure}

To put community and household wealth in our study sample into context, we show where they fit into the Nigerian wealth distribution. Figure \ref{f:hhwealth_dist} shows the distribution of household wealth in the representative DHS sample (black line), and in poor (rich) communities in our study sample, depicted by the blue (gray) line. First, the plot reveals that our study sample does not include the poorest Nigerian households. This is not surprising since Ekiti and Enugu are located in the relatively wealthier Centre-South of the country, with the poorest states located in the North. Second, our sample households are mostly located close to the median of Nigerian household wealth.

\begin{figure}[ht]
	\centering
	\caption{Distribution of study communities in DHS community wealth deciles for Nigeria} \label{f:tuwealth}	\includegraphics[width=.9\textwidth]{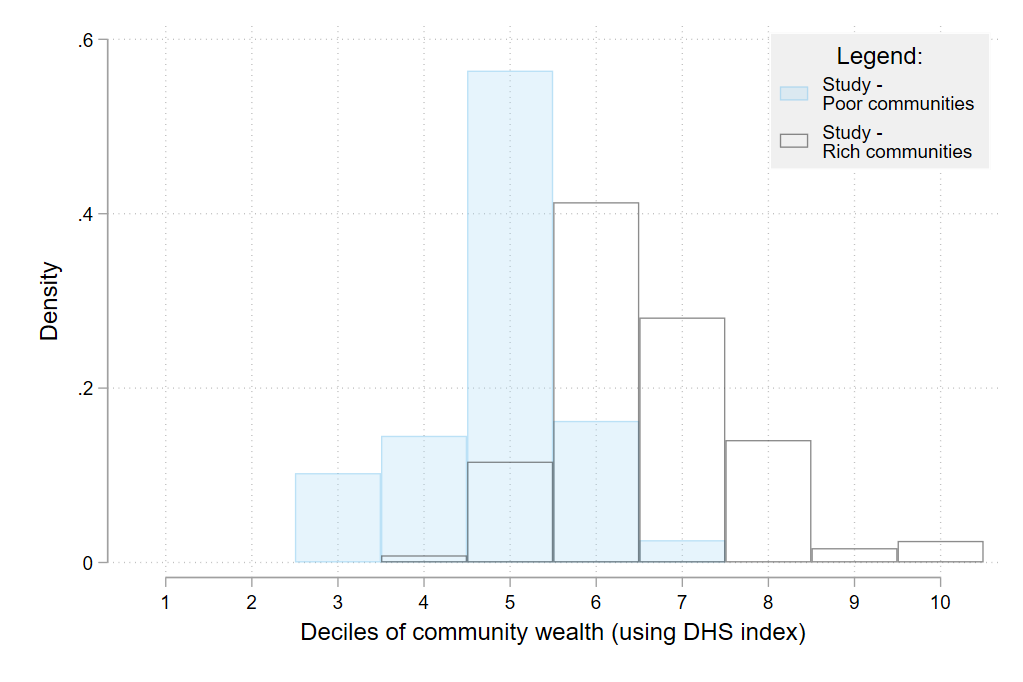}
	\caption*{\emph{Note:} Distribution of the communities in our study along community wealth deciles for the whole of Nigeria, estimated using the 2013 DHS. The wealth index used for this comparison was constructed using a set of questions that was included in both our survey and the DHS household questionnaire.}
\end{figure}

For a similar comparison by community wealth, we construct community wealth deciles for Nigeria, and place poor and rich study communities into these (see Figure \ref{f:tuwealth}).\footnote{ We calculate the median DHS index value within each DHS cluster, and within each study community. We define deciles using only the DHS sample, and use the decile cutoff values to assign study communities to the corresponding wealth deciles. Community wealth is determined using the DHS index to ensure comparability. We retain, however, the classification of rich and poor communities used in the remainder of the paper, which is defined using the more comprehensive Study wealth index.} Our study communities are typically located towards the middle (4th to 7th decile) of the Nigerian community wealth distribution, rather than in the tails.
Some study communities that we have classified as poor (blue), using the median in our study, fall into the higher deciles of the Nigerian community wealth distribution. Overall, however, 80\% of the poor communities in our sample (blue) are in deciles 3 to 5 of the Nigerian distribution, while 88\% of the richest communities in our sample (white) are in the upper deciles of the Nigerian distribution, i.e. deciles 6 to 10. Hence, poor communities in our sample are generally below the median of the national distribution. In contrast, the majority of rich communities in our study are richer than the median community in Nigeria. 

\section{Additional information for international comparison of CLTS impacts}\label{Apsec:comp}

We calculate average night light intensities at the lowest geographical level available in each study. Table \ref{Apt:nightlev} gives a detailed overview of the level of geographical disaggregation and the number of observed geographical units. 

\begin{table}[ht]
    \centering 
	\begin{threeparttable}
    \caption{Summary statistics: night light intensity by study}   \label{Apt:nightlev}
    \footnotesize 
    \begin{tabular}{l l c c c }  
    \toprule
    & \multicolumn{2}{c}{Level of Night Light extraction} & \multicolumn{2}{c}{Summary statistics} \\\cmidrule(lr){2-3}\cmidrule(lr){4-5}
    Country of study & Unit & N  & Mean ($\mu$) & Mean ($\sigma$)  \\
     & (1) & (2)  & (3) & (4)  \\
    \midrule
    Bangladesh & Sub-district & 1 & 0.93 & 2.14  \\
    [1em]
    India & Sub-district & 15 & 1.60 & 2.76  \\
    \hspace{5pt} \emph{Low} &  & 8 & 0.54 & 1.60  \\
    \hspace{5pt} \emph{High} &  & 7 & 2.82 & 4.08  \\
    [1em]
    Indonesia & Cluster & 160 & 7.82 & 4.04  \\
    \hspace{5pt} \emph{Low} &  & 81 & 5.35 & 2.86  \\
    \hspace{5pt} \emph{High} &  & 79 & 10.35 & 5.25 \\
    [1em]
    Mali &  State & 1 & 0.13  & 1.59  \\
    [1em]
    Nigeria & Cluster & 242 & 2.06 & 2.17  \\
    \hspace{5pt} \emph{Low} &  & 120 & 0.33 & 1.01  \\
    \hspace{5pt} \emph{High} &  & 122 & 3.77 & 3.31  \\
    [1em]
    Tanzania & Cluster & 178 & 0.01   & 0.23  \\
    \hspace{5pt} \emph{Low} &  & 167 & 0.00  & 0.24  \\
    \hspace{5pt} \emph{High} &  & 11 & 1.07  & 0.34  \\
    \bottomrule
    \end{tabular}%   
    \begin{tablenotes}[flushleft]
			\item \emph{Notes:} Geographical unit at which night light data was measured in each country of study, and summary statistics about this data. Column 1 shows the lowest administrative unit at which we could map both survey observations and location, in each case. For example, in the case of Bangladesh, all we know is that the study was conducted in the sub-district of Tanore. In the case of India, on the other hand, we managed to map each survey observation to one of 15 sub-districts in the state of Madhya Pradesh in which the study was conducted. Column 2 states how many different units were identified in each study. Column 3 estimates the mean of $\mu$ night light intensity for each country. In the studies in Bangladesh and Mali, this is the mean of a single value: average night light intensity over the single unit available (Sub-district and State, respectively). Column 4 shows the average within-unit standard deviation. Night light intensity is truncated at zero, so higher average intensities are associated with higher within-unit variation. \textit{Low} and \textit{High} subsamples show the split along the within-study median value for night light intensity used for Figure \ref{f:compare}.
		\end{tablenotes}
	\end{threeparttable}
\end{table}

In the Nigerian, Indonesian and Tanzanian datasets, GPS coordinates at the household level are available, so we calculate average night light intensities within a 10km radius of each cluster centroid. In India, we calculate sub-district level measures of night light for 15 sub-districts in which the study was conducted. We exploit within-study variation in night light intensity to calculate treatment effects separately for the poor and richest halves of each sample. The exceptions are Mali and Bangladesh, for whom detailed location information is not available. In Mali (Bangladesh), we assign the average night light intensity of the state (sub-district) in which each study took place, but do not estimate heterogeneous treatment effects.\footnote{The Mali experiment was conducted in the state of Koulikoro, the Bangladesh experiment in the sub-district of Tanore. Column 4 in Table \ref{Apt:nightlev} illustrates the high within-state variation in night light intensity around a very small mean intensity, driven by Koulikoro.}\\

%%%%%%%%%%%%%%%%%%%%%%%%%%%%%%%%%%%%%%
\ignore{
\subsection{Additional sensitivity analysis for the international comparison of CLTS impacts}As additional sensitivity analysis, we once again check whether the cross-context relationship between CLTS effectiveness and SES status might simply be the results of higher treatment effects in areas with lower baseline coverage. To do so, we repeat the analysis we conducted with nightlight intensity with baseline sanitation coverage instead.\footnote{ No data was collected at baseline in the trial conducted in Tanzania, so endline toilet ownership and OD rates among the control group are used instead.}

Figure \ref{f:compare2}, which mirrors Figure \ref{f:compare}, shows that there is no clear relationship between baseline toilet coverage and CLTS treatment effects, suggesting that - in line with our findings for Nigeria specifically - the gradient in CLTS treatment effects along the night light intensity dimension is unlikely driven by pre-existing differences in toilet ownership at baseline across contexts.

\begin{figure}
	\centering
	\caption{CLTS impacts on OD and toilet ownership by average night light intensity} \label{f:compare2}
    \includegraphics[width=\textwidth] {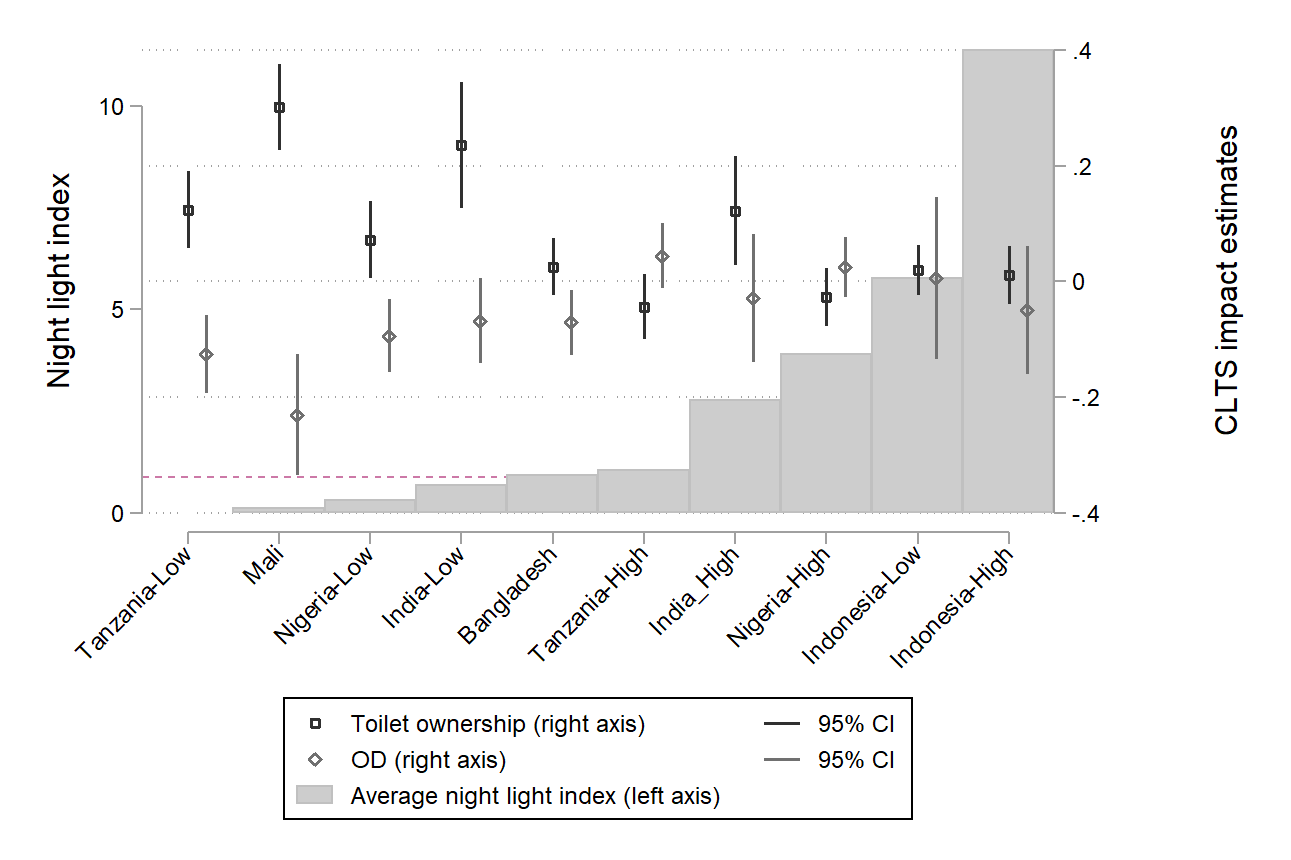}
     %{content/Comparison_onlyCountry.png}
	\caption*{\footnotesize{\emph{Note:} Study-specific point estimates from a simple differences regression of CLTS impacts on OD prevalence (gray squares) and toilet ownership (black dots). Gray bars show the average night light intensity recorded by NASA's Operational Line Scan (OLS) system, on the year of each study's baseline survey, over the study area. Data from the Mali study was unavailable, so results from Table 2 in \cite{pickering2015effect} were used instead (in the case of OD, results for adult women were used).}} 
\end{figure}

We confirm this finding again by plotting the the point estimates from each sample with respect to the logarithm of their baseline toilet ownership coverage rates (Figure \ref{fig:scatter2}). Compared to the Spearman Correlation index with nightlight intensity, the correlation indices for toilet coverage become smaller in magnitude (-0.36 and -0.18) and lose statistical significance, and the explained variation in each plot falls strongly.

\begin{figure}
    \centering
    \caption{CLTS treatment effects by baseline toilet ownership of study area}\label{fig:scatter2}
    \begin{subfigure}[b]{0.45\textwidth}
        \includegraphics[width=\textwidth]{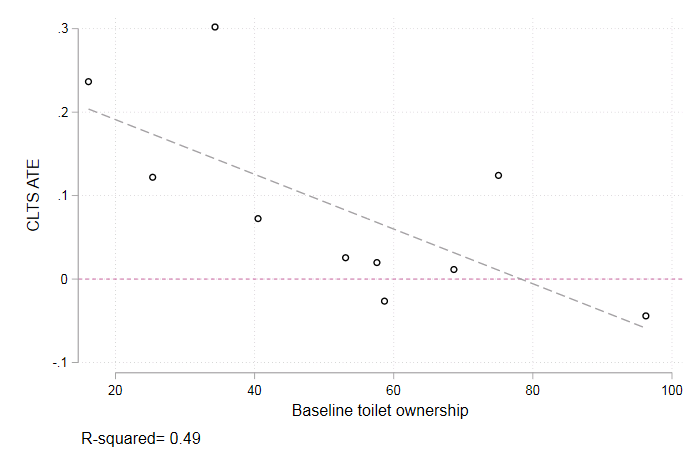}
        \caption{Toilet ownership}
        \label{fig:scatfunclat2}
    \end{subfigure}
    ~ 
    \begin{subfigure}[b]{0.45\textwidth}
        \includegraphics[width=\textwidth]{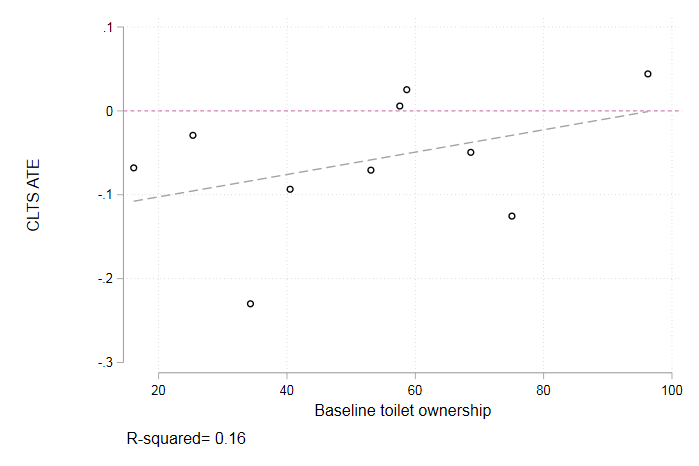}
        \caption{Open defecation}
        \label{fig:scatod2}
    \end{subfigure}
    \caption*{\footnotesize \emph{Note:} Figures plot CLTS average treatment effects on toilet ownership (Panel a) and OD rates (Panel b), from each (sub)sample in countries in Figure \ref{f:compare}, except Mali for which we don't have within country variation. The horizontal axis is the logarithm of night light intensity over each study area during baseline.}
\end{figure}

%%%%%%%%%%%%%%%%%%%%%%%%%%%%%%%%%%%%
As additional complementary evidence, we estimate heterogeneous impact estimates by community wealth within-RCT for those studies where household assets were elicited, namely India, Indonesia and Tanzania. The results in Table \ref{t:tuhte_wealth} show that community wealth-specific impacts are also found in India and Tanzania. While toilet ownership increases and open defecation decreases in both rich and poor communities, CLTS impacts are weakly stronger in poorer communities. Note that the study areas in these countries are on average considerably poorer than those in Nigeria (as measured by nightlight intensity, see Figure XXX in appendix YYY). Indonesia, where the study was implemented in areas that overall look richer than Nigeria, shows small effects throughout.

\begin{table}[ht]
	\centering
	\begin{threeparttable}
		\caption{CLTS impacts by community wealth} \label{t:tuhte_wealth}
		\footnotesize
		\input{content/TU_countries.tex}
		\begin{tablenotes}[flushleft]
			\item \emph{Notes:} All specifications control for XX.
		\end{tablenotes}
	\end{threeparttable}
\end{table}}

\end{appendix}

%% file: content/att1.tex
{
\def\sym#1{\ifmmode^{#1}\else\(^{#1}\)\fi}
\begin{tabular}{l*{3}{c}}
\hline\hline
Completed interviews as \% of total attempted        &\multicolumn{3}{c}{Survey wave}                  \\\cmidrule(lr){2-4}
    &\multicolumn{1}{c}{FU1}&\multicolumn{1}{c}{FU2}&\multicolumn{1}{c}{FU3}\\
                            &\multicolumn{1}{c}{(1)}&\multicolumn{1}{c}{(2)}&\multicolumn{1}{c}{(3)}\\
\hline
Whole sample                &       2.53         &       8.81         &       11.58         \\
[1em]
\emph{By treatment status:} & & & \\
\hspace{5pt} CLTS                        &       2.51         &       8.61         &       12.04         \\
\hspace{5pt} Control                     &       2.55         &       9.01         &       11.14        \\
[1em]
\emph{By community level wealth:} & & & \\
\hspace{5pt} Rich communities            &       3.21         &       9.85         &      14.09         \\
\hspace{5pt} Poor communities            &       1.80         &       7.69         &       8.98        \\
[1em]
\hline\hline
\end{tabular}
}

%% file: content/att2.tex
\begin{tabular}{l*{9}{c}}
\hline\hline
Survey wave:        &\multicolumn{3}{c}{FU1}               &\multicolumn{3}{c}{FU2}               &\multicolumn{3}{c}{FU3}               \\\cmidrule(lr){2-4}\cmidrule(lr){5-7}\cmidrule(lr){8-10}
                    &\multicolumn{1}{c}{(1)}&\multicolumn{1}{c}{(2)}&\multicolumn{1}{c}{(3)}&\multicolumn{1}{c}{(4)}&\multicolumn{1}{c}{(5)}&\multicolumn{1}{c}{(6)}&\multicolumn{1}{c}{(7)}&\multicolumn{1}{c}{(8)}&\multicolumn{1}{c}{(9)}\\
\hline
Treatment Status    &        0.00&       -0.00&        0.00&        0.01&       -0.01&       -0.01&        0.01&        0.00&        0.01\\
                    &      (0.90)&      (0.96)&      (0.90)&      (0.60)&      (0.49)&      (0.48)&      (0.22)&      (0.71)&      (0.55)\\
[1em]
Poor community      &            &            &       -0.01&            &            &        0.00&            &            &       -0.03\\
                    &            &            &      (0.06)&            &            &      (0.74)&            &            &      (0.02)\\
\hline
Household controls  &          No&         Yes&         Yes&          No&         Yes&         Yes&          No&         Yes&         Yes\\
Number of observations&       4,647&       4,505&       4,505&       4,647&       4,505&       4,505&       4,722&       4,546&       4,546\\
\hline\hline
\end{tabular}

%% file: content/table_implementation_lock.tex
\begin{tabular}{l*{4}{c}}
\hline\hline
					&\multicolumn{4}{c}{Dep.variable: main respondent performs OD}       \\\cmidrule(lr){2-5}
					&\multicolumn{1}{c}{By state}&\multicolumn{3}{c}{IV approach}       \\\cmidrule(lr){2-2}\cmidrule(lr){3-5}
Sample:             &\multicolumn{1}{c}{All}&\multicolumn{1}{c}{All}&\multicolumn{1}{c}{Rich communities}&\multicolumn{1}{c}{Poor communities}\\
                    &\multicolumn{1}{c}{(1)}&\multicolumn{1}{c}{(2)}&\multicolumn{1}{c}{(3)}&\multicolumn{1}{c}{(4)}\\
\hline
CLTS x Enugu    	&       -0.02&            &            &            \\
                    &      (0.43)&            &            &            \\
[1em]
CLTS x Ekiti 		&       -0.05&            &            &            \\
p-val               &      (0.04)&            &            &            \\
[1em]
Difference          &       -0.03&            &            &            \\
                    &      (0.33)&            &            &            \\
[1em]
CLTS        		&            &       -0.04&        0.02&       -0.10\\
                    &            &      (0.04)&      (0.55)&      (0.00)\\
\hline
First-stage F-statistic	&            &      645.50&      174.67&      817.17\\
No. of communities  	&         246&       246     &     123       &        123    \\
No. of observations &      12,697&      12,697&       6,515&       6,182\\
\hline\hline
\end{tabular}

%% file: content/tableTUHTE_OD_2_CONT.tex
\begin{tabular}{l*{5}{c}}
\hline\hline
					&\multicolumn{5}{c}{Dep. variable: main respondent performs OD} \\\cmidrule(lr){2-6}
Community Characteristic (CC) at BL:   &Asset wealth&Night lights&Density&Isolation&Toilet coverage\\
                    &\multicolumn{1}{c}{(1)}&\multicolumn{1}{c}{(2)}&\multicolumn{1}{c}{(3)}&\multicolumn{1}{c}{(4)}&\multicolumn{1}{c}{(5)}\\
\hline
CLTS     				 					&       -0.04&       -0.03&       -0.04&       -0.04&       -0.03\\
\hspace{5pt} \emph{p}-value (naive)			&      (0.04)&      (0.05)&      (0.04)&      (0.04)&      (0.04)\\
\hspace{5pt} \emph{p}-value (MHT robust)	&      [0.07]&      [0.07]&      [0.07]&      [0.07]&      [0.07]   \\
[1em]
CLTS x CC		             				&        0.06&        0.00&        0.04&       -0.04&        0.03\\
\hspace{5pt} \emph{p}-value (naive)			&      (0.00)&      (0.98)&      (0.02)&      (0.05)&      (0.15)\\
\hspace{5pt} \emph{p}-value (MHT robust)	&      [0.01]&      [0.98]&      [0.09]&      [0.22]&      [0.53]   \\
\hline
No. of communities          				&         246   &         246   &         246   &         246   &         246   \\
No. of observations  						&      12,697   &      12,697   &      12,697   &      12,697   &      12,697   \\
\hline\hline
\end{tabular}

%% file: content/tableSESrobust.tex
\begin{tabular}{l*{4}{c}}
\hline\hline
Asset wealth compared to:&Night lights&     Density&   Isolation&         All\\
                    &\multicolumn{1}{c}{(1)}&\multicolumn{1}{c}{(2)}&\multicolumn{1}{c}{(3)}&\multicolumn{1}{c}{(4)}\\
\hline
CLTS                &        0.04&        0.05&        0.03&        0.05\\
                    &      (0.11)&      (0.07)&      (0.15)&      (0.05)\\
[1em]
Low asset wealth    &        0.08&        0.08&        0.08&        0.08\\
                    &      (0.00)&      (0.00)&      (0.00)&      (0.00)\\
[1em]
CLTS x Low asset wealth&       -0.09&       -0.09&       -0.09&       -0.09\\
                    &      (0.01)&      (0.01)&      (0.01)&      (0.01)\\
[1em]
Low night lights    &        0.07&            &            &        0.04\\
                    &      (0.00)&            &            &      (0.12)\\
[1em]
CLTS x Low night lights&       -0.05&            &            &       -0.03\\
                    &      (0.10)&            &            &      (0.55)\\
[1em]
Low density         &            &        0.08&            &        0.06\\
                    &            &      (0.00)&            &      (0.03)\\
[1em]
CLTS x Low density  &            &       -0.08&            &       -0.07\\
                    &            &      (0.01)&            &      (0.07)\\
[1em]
High isolation      &            &            &        0.07&        0.01\\
                    &            &            &      (0.00)&      (0.78)\\
[1em]
CLTS x High isolation&            &            &       -0.05&        0.01\\
                    &            &            &      (0.09)&      (0.91)\\
\hline
No. of observations &      12,697&      12,697&      12,697&      12,697\\
\hline\hline
\end{tabular}

%% file: content/table_wealth_lock.tex
{
\def\sym#1{\ifmmode^{#1}\else\(^{#1}\)\fi}
\begin{tabular}{l*{3}{c}} 
\hline\hline
                    &\multicolumn{3}{c}{Dep. variable: main respondent performs OD}                         \\\cmidrule(lr){2-4}
Sample: &\multicolumn{1}{c}{All communities}&\multicolumn{1}{c}{Rich communities}&\multicolumn{1}{c}{Poor communities}\\
 &\multicolumn{1}{c}{(1)}&\multicolumn{1}{c}{(2)}&\multicolumn{1}{c}{(3)}\\
\hline
CLTS x Rich ($\gamma_r$)                &       -0.01         &        0.01         &       -0.06  \\
p-value (naive)                         &      (0.61)         &      (0.51)         &      (0.07)  \\
p-value (MHT robust)                    &      [0.67]         &      [0.67]         &      [0.21]  \\
[1em]
CLTS x Poor ($\gamma_r + \gamma_d$)     &       -0.06         &        0.01         &       -0.11         \\
p-value (naive)                         &      (0.01)         &      (0.78)         &      (0.00)    \\
p-value (MHT robust)                    &      [0.05]         &      [0.78]         &       [0.01] \\
[1em]
Difference ($\gamma_d$)                 &       -0.05         &       -0.01         &       -0.05         \\
p-value (naive)                         &      (0.04)         &      (0.86)         &      (0.14)         \\
p-value (MHT robust)                    &      [0.18]         &       [0.86]        &       [0.45] \\
\hline
DV control mean (EL, Rich)           &        0.37         &        0.33         &        0.47         \\
DV control mean (EL, Poor)           &        0.60         &        0.45         &        0.69         \\
No. of TUs                   &         246         &         123         &         123         \\
No. of obs.                  &      12,697         &       6,515         &       6,182         \\
\hline\hline
\end{tabular}
}

%% file: content/tableHHHTE2_OD_lock.tex
\begin{tabular}{l*{3}{c}} 
\hline\hline
    &\multicolumn{3}{c}{Dep. variable: main respondent performs OD} \\\cmidrule(lr){2-4}
Household characteristic (at BL): 			&\multicolumn{1}{c}{Low ed. HoH}&\multicolumn{1}{c}{Female HoH}&\multicolumn{1}{c}{Any children}\\
                    						&\multicolumn{1}{c}{(1)}&\multicolumn{1}{c}{(2)}&\multicolumn{1}{c}{(3)}\\
\hline
CLTS x No ($\gamma_{r}$)                    &       -0.04   &       -0.03   &       -0.03   \\
\hspace{5pt} \emph{p}-value (naive)         &      (0.04)   &      (0.13)   &      (0.12)   \\
\hspace{5pt} \emph{p}-value (MHT robust)    &      [0.11]   &      [0.30]   &      [0.30]   \\
[1em]
CLTS x Yes ($\gamma_{r} + \gamma_{d}$)      &       -0.05   &       -0.04   &       -0.05   \\
\hspace{5pt} \emph{p}-value (naive)         &      (0.03)   &      (0.04)   &      (0.03)   \\
\hspace{5pt} \emph{p}-value (MHT robust)    &      [0.08]   &      [0.08]   &      [0.08]   \\
[1em]
Difference ($\gamma_{d}$)                   &       -0.01   &       -0.01   &       -0.02   \\
\hspace{5pt} \emph{p}-value (naive)         &      (0.12)   &      (0.50)   &      (0.37)   \\
\hspace{5pt} \emph{p}-value (MHT robust)    &      [0.45]   &      [0.77]   &      [0.77]   \\
[1em]
\hline
DV control mean (EL, No)&        0.64   &        0.61   &        0.63   \\
DV control mean (EL, Yes)&        0.69   &        0.65   &        0.61   \\
No. of TUs          &         246   &         246   &         246   \\
No. of obs.         &      12,697   &      12,697   &      12,697   \\
\hline\hline
\end{tabular}

%% file: content/table_channels_TUnlL_lock.tex
{
\def\sym#1{\ifmmode^{#1}\else\(^{#1}\)\fi}
\begin{tabular}{l*{4}{c}}
\hline\hline
Outcome =1 if:      & Owns toilet         &Owns functioning toilet         &Usage (if functioning)         &  Shared use         \\
                    &\multicolumn{1}{c}{(1)}         &\multicolumn{1}{c}{(2)}         &\multicolumn{1}{c}{(3)}         &\multicolumn{1}{c}{(4)}         \\
\hline
\\
\multicolumn{5}{c}{\emph{Heterogeneous impacts by nightlight intensity}}\\

CLTS x High ($\gamma_{r}$)                  &       -0.01       &       -0.02       &       -0.02       &        0.02  \\
\hspace{5pt} \emph{p}-value (naive)         &      (0.70)       &      (0.45)       &      (0.31)       &      (0.02)         \\
\hspace{5pt} \emph{p}-value (MHT robust)    &      [0.70]       &      [0.57]       &      [0.57]       &      [0.05]         \\
[1em]
CLTS x Low ($\gamma_{r} + \gamma_{d}$)      &        0.05       &        0.08       &        0.03       &        0.00         \\
\hspace{5pt} \emph{p}-value (naive)         &        0.05       &        0.01       &        0.13       &        0.97         \\
\hspace{5pt} \emph{p}-value (MHT robust)    &      [0.18]       &      [0.01]       &      [0.25]       &      [0.97]         \\
[1em]
Difference ($\gamma_{d}$)                   &        0.06       &        0.09       &        0.05       &       -0.01         \\
\hspace{5pt} \emph{p}-value (naive)         &      (0.10)       &      (0.01)       &      (0.08)       &      (0.36)         \\
\hspace{5pt} \emph{p}-value (MHT robust)    &      [0.19]       &      [0.02]       &      [0.19]       &      [0.36]         \\
\hline
DV control mean (EL, High)&        0.45&        0.45&        0.57&        0.03\\
DV control mean (EL, Low)&        0.29&        0.28&        0.48&        0.03\\
\hline
\\
\multicolumn{5}{c}{\emph{Heterogeneous impacts by household density}}\\

CLTS x High ($\gamma_{r}$)                  &       -0.02   &       -0.01&       -0.00&        0.01\\
\hspace{5pt} \emph{p}-value (naive)         &      (0.30)   &      (0.54)&      (0.80)&      (0.03)\\
\hspace{5pt} \emph{p}-value (MHT robust)    &      [0.56]   &      [0.78]&      [0.80]&      [0.09]         \\
[1em]
CLTS x Low ($\gamma_{r} + \gamma_{d}$)      &        0.07   &        0.09&        0.02&        0.00\\
\hspace{5pt} \emph{p}-value (naive)         &        0.01   &        0.00&        0.43&        0.81\\
\hspace{5pt} \emph{p}-value (MHT robust)    &      [0.03]   &      [0.01]&      [0.69]&      [0.81]         \\
[1em]
Difference ($\gamma_{d}$)                   &        0.09   &        0.10&        0.02&       -0.01\\
\hspace{5pt} \emph{p}-value (naive)         &      (0.01)   &      (0.01)&      (0.43)&      (0.63)\\
\hspace{5pt} \emph{p}-value (MHT robust)    &      [0.03]   &      [0.02]&      [0.68]&      [0.68]         \\
\hline
DV control mean (EL, High)&        0.46&        0.45&        0.58&        0.03\\
DV control mean (EL, Low)&        0.26&        0.25&        0.44&        0.03\\
\hline
\\
\multicolumn{5}{c}{\emph{Heterogeneous impacts by degree of isolation}}\\

CLTS x High ($\gamma_{r}$)                  &        0.05       &        0.08         &        0.02         &        0.00         \\
\hspace{5pt} \emph{p}-value (naive)         &      (0.05)       &      (0.01)         &      (0.34)         &      (0.77)         \\
\hspace{5pt} \emph{p}-value (MHT robust)    &      [0.12]       &      [0.01]         &      [0.59]         &      [0.77]         \\
[1em]
CLTS x Low ($\gamma_{r} + \gamma_{d}$)      &       -0.01         &       -0.01         &       -0.01         &        0.01         \\
\hspace{5pt} \emph{p}-value (naive)         &        0.79         &        0.68         &        0.64         &        0.08         \\
\hspace{5pt} \emph{p}-value (MHT robust)    &      [0.90]        &      [0.90]         &      [0.90]         &      [0.23]         \\
[1em]
Difference ($\gamma_{d}$)                   &       -0.06         &       -0.09        &       -0.03         &        0.01         \\
\hspace{5pt} \emph{p}-value (naive)         &      (0.08)         &      (0.01)         &      (0.32)         &      (0.68)         \\
\hspace{5pt} \emph{p}-value (MHT robust)    &      [0.22]        &      [0.03]         &      [0.54]         &      [0.68]         \\
\hline
DV control mean (EL, High)&        0.30&        0.29&        0.52&        0.03\\
DV control mean (EL, Low)&        0.43&        0.42&        0.54&        0.03\\
\hline
Number of communities&         246         &         246         &         245         &         246         \\
Number of observations&      12,497         &      12,497         &       7,113         &      12,697         \\
\hline\hline
\end{tabular}
}

%% file: content/TU_countries.tex
\begin{tabular}{l*{6}{c}}
\hline\hline
                    &\multicolumn{3}{c}{Functioning Latrine}&\multicolumn{3}{c}{Open Defecation}   \\\cmidrule(lr){2-4}\cmidrule(lr){5-7}
                    &\multicolumn{1}{c}{(1)}&\multicolumn{1}{c}{(2)}&\multicolumn{1}{c}{(3)}&\multicolumn{1}{c}{(4)}&\multicolumn{1}{c}{(5)}&\multicolumn{1}{c}{(6)}\\
                    &\multicolumn{1}{c}{India}&\multicolumn{1}{c}{Indonesia}&\multicolumn{1}{c}{Tanzania}&\multicolumn{1}{c}{India}&\multicolumn{1}{c}{Indonesia}&\multicolumn{1}{c}{Tanzania}\\
\hline
treat\_poor          &        0.13&       -0.02&        0.08&       -0.08&       -0.02&       -0.01\\
                    &      (0.03)&      (0.52)&      (0.20)&      (0.21)&      (0.80)&      (0.83)\\
[1em]
treat               &        0.12&        0.02&        0.08&       -0.08&       -0.01&       -0.05\\
                    &      (0.01)&      (0.34)&      (0.06)&      (0.05)&      (0.88)&      (0.33)\\
\hline
No. of obs.         &     1687.00&     1908.00&     1773.00&     1773.00&     1908.00&     1687.00\\
\hline\hline
\end{tabular}